\documentclass[twocolumn,showpacs,preprintnumbers,amsmath,amssymb]{revtex4}

\usepackage{graphicx}
\usepackage{dcolumn}
\usepackage{bm}

\begin{document}

\title{Scattering function for a self-avoiding polymer chain}

\author{A.D. Drozdov}
\affiliation{Department of Chemical Engineering\\
Ben-Gurion University of the Negev\\
P.O. Box 653,
Beer-Sheva 84105, Israel.}
\email{aleksey@bgumail.bgu.ac.il}

\date{\today}

\begin{abstract}
An explicit expression is derived for the scattering
function of a self-avoiding polymer chain in a
$d$-dimensional space.
The effect of strength of segment interactions on the
shape of the scattering function and the radius of gyration
of the chain is studied numerically.
Good agreement is demonstrated between experimental
data on dilute solutions of several polymers
and results of numerical simulation.
\end{abstract}

\pacs{31.15.Kb, 61.12.Bt, 82.35.Lr}
\maketitle

\section{Introduction}

This paper is concerned with the scattering function
(form factor) of a self-repellent polymer chain.
The effect of excluded-volume interactions between segments
on the characteristic size of a ``real" chain
has attracted substantial attention from the beginning of
the 1950s \cite{Flo49,Bue53,Fix55}.
Although a noticeable progress has been reached in this area
(the results of investigation are summarized
in monographs
\cite{dGen79,DE86,CJ90,Yam97,Sch99}),
the effect of self-avoiding interactions on the
statistics of macromolecules has remained a subject
of debate in the past decade,
see \cite{DTC00,KSS01,PR00,AK02,Lam02,BW03,CG04,KM04,LLH04},
to mention a few.
This fact may be ascribed to the importance of
excluded-volume interactions for the description of
such phenomena as
(i) protein folding and denaturation \cite{KSS01,CG04},
(ii) unzipping of DNA \cite{LLH04},
(iii) mechanical response of individual macromolecules \cite{Lam02},
and (iv) transport of chains through
narrow pores \cite{KM04}.
A correct account for self-repellent interactions
(i) provides a consistent description of
the effect of molecular weight on intrinsic viscosity of
dilute polymer solutions \cite{KTE96},
(ii) leads to an adequate characterization
of time-dependent shear stresses
\cite{PR00,BW03} and transport coefficients in polymer
fluids \cite{AK02},
(iii) predicts reinforcement of interfaces between
incompatible solvents by random copolymers \cite{DTC00}, etc.

The influence of self-avoiding interactions
on the scattering intensity of a chain in light,
X-ray, and neutron scattering experiments has
been investigated in
\cite{HRW58,SB68,MRF97,ANN98,PHB00,FAB02,Gol03,PS04},
where phenomenological equations have been developed for
the structure function based on simplified theoretical models
and results of Monte Carlo simulation.
Although these relations improve our knowledge of the
statistics of self-avoiding chains, their applicability
remains questionable for the lack of direct comparison
of their predictions with those grounded on an exact
expression for the form factor.
The objective of this study is to derive analytical
formulas for the scattering function of a self-avoiding chain
in a $d$-dimensional space and its radius of gyration
and to analyze the influence of strength of excluded-volume
interactions on these quantities.

Two approaches are conventionally employed to model
a self-repellent flexible chain.
According to the first, a chain is treated as a set
of freely-jointed rigid segments linked in sequel,
whereas self-avoiding interactions do not permit each
pair of segments to occupy the same place \cite{Flo49}.
This model implies that the distribution function of
end-to-end vectors for the chain coincides with that for a
self-avoiding random walk on a lattice \cite{MS93,Law96}.
Although the latter function is unknown,
it is believed that the leading term in the
expression for the distribution function of a long chain
is described by a stretched exponential function multiplied
by a polynomial of the end-to-end distance \cite{MM71,Clo74}.

The other approach (the Edwards model \cite{Edw65})
treats an arbitrary configuration of a chain
as a curve with a Hamiltonian ascribed to it.
The Hamiltonian equals the sum of the conventional
Hamiltonian for a Gaussian chain and the excluded-volume
functional, where the energy of segment interactions
is approximated by the delta-function.
This method does not exclude intersections between
segments, but assumes their number to be reduced
(compared to that for a Gaussian chain) in some proportion
to the strength of self-repellent interactions.
The distribution function of end-to-end vectors for a
self-avoiding chain is determined
by the conventional path integral (Green's function),
and the scattering function (form factor) is given by
the classical formula \cite{HB94}, where averaging
is performed with the help of the Green function.

Despite an apparent simplicity of the Edwards model,
no exact formulas have yet been derived
within this approach for the radial distribution function
and the form factor.
This may be explained by the fact that the Hamiltonian
containing the excluded-volume functional becomes non-quadratic,
and the path integral cannot be computed by using
conventional approaches \cite{Kle95}.
Among non-rigorous techniques proposed for the treatment of
this problem, it is worth mentioning perturbative
\cite{Fix55,FF66,YT67,MN84}
and renormalization-group \cite{OO82,ASD01,SE04} methods.

The purpose of the present paper is to derive an exact
formula for the form factor of a self-avoiding chain
in a $d$-dimensional space based of the Edwards model
with a slight modification of the excluded-volume functional.
This modification is caused by the observation that the average
(over configurations) value of the Edwards Hamiltonian,
as well as all moments of the number of segment intersections
become infinite at $d>1$.
The divergence reflects an increase in the number of intersections
between different segments of the chain when the distance between
them (measured along the curve that describes an arbitrary configuration)
approaches zero.
Based on a similarity between configurations of a
Gaussian chain and trajectories of a standard Brownian
motion, the latter may be explained by the well-known
properties of the local time of intersection for a
$d$-dimensional Brownian path \cite{KS91}.

We concentrate on the derivation of an explicit expression
for the scattering function.
The interest to this quantity is explained by the fact that
a number of experimental data on small-angle
X-ray and neutron scattering by macromolecules is available,
which allow the strength of excluded-volume interactions
to be found by fitting observations.
An analytical formula for the distribution function
of end-to-end vectors will be developed in a subsequent paper.
The exposition is organized as follows.
The average number of self-intersections for a flexible
chain is calculated in Section 2.
The Edwards model with a regularized Hamiltonian
is introduced in Section 3.
A series expansion for the scattering function is developed
in Section 4.
Borel's summation of the series is performed in Section 5,
where a closed-form expression is derived for the Laplace
transform of this function.
Explicit expressions for the scattering function of a self-repellent
chain are obtained in Section 6 for the one-dimensional case
and in Section 7 for the three-dimensional case.
Two asymptotic approaches are suggested for the simplification
of the form factor for a self-avoiding chain.
The approximate expressions are tested in Section 8,
where results of numerical simulation are compared with
experimental data.
Some concluding remarks are formulated in Section 9.

\section{The local time of self-intersections}

We consider a standard Gaussian chain in a $d$-dimensional
space.
One end of the chain is fixed at the origin,
and the other end is located at a point with radius vector
${\bf Q}$.
An arbitrary configuration of the chain is described by
a curve ${\bf r}(\tau)$ with $\tau\in [0,L]$,
where $L$ stands for the contour length.
The Hamiltonian of the chain reads
\begin{equation}\label{1}
H_{0}({\bf r})=\frac{dk_{\rm B}T}{2b_{0}}\int_{0}^{L}
\Bigl (\frac{{\rm d}{\bf r}}{{\rm d}\tau}(\tau)\Bigr )^{2}
{\rm d}\tau ,
\end{equation}
where $b_{0}$ denotes the length of a statistical
segment,
$k_{\rm B}$ is Boltzmann's constant,
and $T$ is the absolute temperature.

Fixing a segment of the chain labeled by $t\in (0,L]$,
we consider intersections of this segment with other
segments labeled by $\tau \in (s_{1},s_{2})$, where
$0\leq s_{1}<s_{2}<t$.
The number of intersections (the local time of intersection)
is defined as
\begin{equation}\label{2}
\nu(s_{1},s_{2};t)
= b_{0}^{d-1} \int_{s_{1}}^{s_{2}}
\delta({\bf r}(t)-{\bf r}(\tau)) {\rm d} \tau ,
\end{equation}
where
\begin{equation}\label{3}
\delta({\bf r})=\frac{1}{(2\pi)^{d}}
\int \exp (-\imath {\bf k}\cdot {\bf r})
{\rm d}{\bf k}
\end{equation}
is the Dirac function,
the dot stands for inner product,
and integration is performed over all vectors ${\bf k}$.
The delta-function in Eq. (\ref{2}) counts the ``number" of
intersections between the segments labeled by $t$ and $\tau$,
and the integral describes the sum these ``numbers"
over various segments $\tau$ belonging
to the interval $(s_{1},s_{2})$.
The pre-factor $b_{0}$ is introduced
to make the right-hand side of Eq. (\ref{2}) dimensionless.
The dependence of $\nu$ on the end-to-end vector ${\bf Q}$
is suppressed.

Although Eq. (\ref{2}) is widely used,
it should be treated with caution
for this formula has an unambiguous meaning when
the average (over configurations) number of self-intersections
$\langle \nu(s_{1},s_{2};t)\rangle_{\bf Q}$ is finite
for any point $t$ and an arbitrary interval $(s_{1},s_{2})$.
This is not the case, however, because all moments of the function
$\nu(s_{1},s_{2};t)$ diverge as $s_{2}\to t$.
This fact appears to be natural, for all configurations are
taken into account in the calculation of the average value
of $\nu$, including those curves ${\bf r}(\tau)$ that
``oscillate" (and intersect themselves) with high frequencies.

Several approaches were developed to avoid the divergence
of the integral in Eq. (\ref{2}).
The simplest method \cite{Yor85}
consists in (i) replacement of the delta-function
in Eq. (\ref{2}) with another delta-function that
``measures" the number of intersections ``shifted" to
some vector ${\bf a}$,
(ii) determination of the average number of intersections,
and (iii) calculation of its limit (if it exists) as
${\bf a}\to {\bf 0}$.
This technique allows the first moment of the number of
intersections to be found, but does not permit
the entire set of moments to be determined analytically.

The renormalization group approach was applied
to the analysis of intersections of Brownian motion
in \cite{Ros86,Dyn88,FDS00}.
An appropriate renormalization (both multiplicative and additive)
allows (i) explicit expressions for the first two moments of
$\nu$ to be derived,
and (ii) convergence of the ``renormalized"
process $\nu(s_{1},s_{2},t)$
[treated as a function of $t$ for a given interval $(s_{1},s_{2})$]
to a Brownian motion with known characteristics to be proved.
Although these methods seem rather promising,
their interpretation in terms of physical
parameters of a chain remains unclear.

To avoid complications of the renormalization-group approach,
we regularize the number of intersections by ascribing to each
intersection between segments labeled by $\tau$ and $t$
some weight $\rho$
that (i) approaches zero when the difference $t-\tau$ vanishes,
and (ii) reaches its ultimate value of unity when
the ``distance" between these points
along the curve ${\bf r}(\tau)$ becomes relatively large.
This means that we replace definition (\ref{2}) with
\begin{equation}\label{4}
\nu(s_{1},s_{2};t)=b_{0}^{d-1} \int_{s_{1}}^{s_{2}}
\rho(t-\tau) \delta ({\bf r}(t)-{\bf r}(\tau)) {\rm d}\tau.
\end{equation}
As the regularizing function $\rho(t)$
may be chosen from a set of functions satisfying the conditions
$\rho(0)=0$ and $\lim_{t\to \pm \infty} \rho(t)=1$,
Eq. (\ref{4}) provides an additional degree of freedom in the
design of models for self-avoiding chains.
In what follows, we set
\begin{equation}\label{5}
\rho(t)=1-\exp (|t|/L_{0}),
\end{equation}
where $L_{0}$ is the characteristic length of internal
inhomogeneity.
Roughly speaking, Eqs. (\ref{4}) and (\ref{5}) treat intersections
between segments located [along the curve ${\bf r}(\tau)$]
at distances exceeding $L_{0}$
with the same (unit) weight and disregard
intersections between segments located [along this curve]
closer than the characteristic length $L_{0}$.

To find the average number of self-intersections,
we replace the delta-function in Eq. (\ref{4})
with its Fourier transform (\ref{3}), and
average the obtained relation over all configuration of a
chain with an end-to-end vector ${\bf Q}$,
\begin{eqnarray*}
\langle \nu(s_{1},s_{2};t)\rangle_{\bf Q}
&=& \frac{b_{0}^{d-1}}{(2\pi)^{d}}
\int_{s_{1}}^{s_{2}} \rho(t-\tau) {\rm d}\tau
\nonumber\\
&&\hspace*{-24 mm}\times
\int \Bigl \langle \exp \Bigl [-\imath {\bf k}
\cdot \Bigl ({\bf r}(t)-{\bf r}(\tau)\Bigr )
\Bigr ] \Bigr \rangle_{\bf Q} {\rm d}{\bf k}
\nonumber\\
&=& \frac{b_{0}^{d-1}}{(2\pi)^{d}}
\int_{s_{1}}^{s_{2}} \rho(t-\tau) {\rm d}\tau
\nonumber\\
&&\hspace*{-23 mm}\times
\int \exp \Bigl (-\frac{b_{0}k^{2}}{2d} \Delta (t,\tau)
-\imath {\bf k}\cdot {\bf Q}
\frac{t-\tau}{L} \Bigr ) {\rm d}{\bf k},
\end{eqnarray*}
where the function $\Delta(t,\tau)$ describes correlations
between segments.
Calculation of the Gaussian integral results in
\begin{eqnarray}\label{6}
\langle \nu(s_{1},s_{2};t)\rangle_{\bf Q}
&=& \frac{1}{b_{0}} \Bigl (\frac{d b_{0}}{2\pi}\Bigr )^{\frac{d}{2}}
\int_{s_{1}}^{s_{2}}
\frac{\rho(t-\tau)}{\Delta^{\frac{d}{2}}(t,\tau)}
\nonumber\\
&&\times
\exp \Bigl [-\frac{d Q^{2}(t-\tau)^{2}}{2b_{0}L^{2}\Delta(t,\tau)}
\Bigr ] {\rm d}\tau .
\end{eqnarray}
Equation (\ref{6}) describes the average number of
self-intersections for a chain with an arbitrary correlation
function $\Delta(t,\tau)$.
For a standard Gaussian chain with
\[
\Delta(t,\tau)=(t-\tau)\Bigl (1-\frac{t-\tau}{L}\Bigr ),
\]
we arrive at the formula
\begin{eqnarray}\label{7}
\hspace*{-5 mm}&&\langle \nu(s_{1},s_{2};t)\rangle_{\bf Q}
= \frac{1}{b_{0}} \Bigl (\frac{d b_{0}}{2\pi}\Bigr )^{\frac{d}{2}}
\int_{s_{1}}^{s_{2}}
\rho(t-\tau) (t-\tau)^{-\frac{d}{2}}
\nonumber\\
\hspace*{-5 mm}&&\times
\Bigl (1-\frac{t-\tau}{L}\Bigr )^{-\frac{d}{2}}
\exp \Bigl [-\frac{d Q^{2}(t-\tau)}{2b_{0}L^{2}(1-(t-\tau)/L)}
\Bigr ] {\rm d}\tau .
\end{eqnarray}
The average (over chains with various end-to-end vectors ${\bf Q}$)
number of intersections is given by
\begin{equation}\label{8}
\langle \nu(s_{1},s_{2};t)\rangle
=\int \langle \nu(s_{1},s_{2};t)\rangle_{\bf Q}
p_{0}({\bf Q}) {\rm d} {\bf Q},
\end{equation}
where
\begin{equation}\label{9}
p_{0}({\bf Q})=\Bigl (\frac{d}{2\pi b_{0}L}\Bigr )^{\frac{d}{2}}
\exp \Bigl (-\frac{d Q^{2}}{2b_{0}L}\Bigr ).
\end{equation}
is the distribution function of end-to-end vectors
for a Gaussian chain in a $d$-dimensional space.
Insertion of expressions (\ref{7}) and (\ref{9})
into Eq. (\ref{8}) and calculation of the Gaussian integral
yield
\begin{equation}\label{10}
\langle \nu(s_{1},s_{2};t)\rangle
= \frac{1}{b_{0}}\Bigl (\frac{d b_{0}}{2\pi}\Bigr )^{\frac{d}{2}}
\int_{s_{1}}^{s_{2}}
\frac{\rho(t-\tau)}{(t-\tau)^{\frac{d}{2}}} {\rm d}\tau.
\end{equation}
Equation (\ref{10}) shows that the average number of
intersections between a segment labeled by $t$ and
segments belonging to the interval $(s_{1},s_{2})$
approaches infinity when $s_{2}\to t$ and $d>1$,
if a regularization function $\rho(t)$ is not used.

Setting $s_{1}=0$ and $s_{2}=t$ in Eq. (\ref{4}),
we find the regularized number of intersections between
a segment labeled by $t$ and all segments labeled
by $\tau\leq t$.
Summation of these numbers over $t\in (0,L]$
provides us (after division by $L$) with the number of
self-intersections per unit length
\begin{equation}\label{11}
\nu^{\ast}= \frac{1}{L} \int_{0}^{L} \nu(0,t;t){\rm d}t.
\end{equation}
Substitution of expression (\ref{10}) into Eq. (\ref{11})
results in the average number of self-intersections
per unit length of a standard Gaussian chain in a $d$-dimensional
space
\begin{eqnarray}\label{12}
\langle \nu^{\ast}\rangle &=& \frac{1}{b_{0}L}
\Bigl (\frac{d b_{0}}{2\pi}\Bigr )^{\frac{d}{2}}
\int_{0}^{L} {\rm d}t \int_{0}^{t}
\frac{\rho(t-\tau)}{(t-\tau)^{\frac{d}{2}}} {\rm d}\tau
\nonumber\\
&=&
\frac{L}{b_{0}} \Bigl (\frac{d b_{0}}{2\pi L}\Bigr )^{\frac{d}{2}}
\int_{0}^{1} (1-\tau) \rho(L\tau) \tau^{-\frac{d}{2}}{\rm d}\tau ,
\end{eqnarray}
where we set $s=t-\tau$ in the internal integral,
changed the order of integration,
computed the integral over $t$,
and set $\tau=s/L$ in the result.
For a Gaussian chain in a three-dimensional space,
Eqs. (\ref{5}) and (\ref{12}) yield
\[
\langle \nu^{\ast} \rangle = \frac{3}{\pi}
\sqrt{\frac{3b_{0}}{2\pi L}}\Bigl [
\frac{1}{\sqrt{\kappa}}\Bigl (
(\kappa+1) \gamma\Bigl (\frac{1}{2}, \kappa\Bigr )
-\gamma\Bigl (\frac{3}{2},\kappa\Bigr )\Bigr ) -2\Bigr ],
\]
where
\begin{equation}\label{13}
\kappa=\frac{L}{L_{0}},
\qquad
\gamma(a,x)=\int_{0}^{x} z^{a-1}\exp (-z) {\rm d}z .
\end{equation}
When $\kappa\to \infty$, which means that
the contour length of a chain substantially exceeds
the characteristic length of regularization,
the incomplete gamma-function $\gamma(a,\kappa)$
approaches the gamma-function $\Gamma(a)$,
and the leading term in the expression
for the average number of self-intersections
per unit length reads
\begin{equation}\label{14}
\langle \nu^{\ast} \rangle
= \frac{3}{\pi}\sqrt{\frac{3b_{0}}{2L_{0}}}
\qquad
(d=3).
\end{equation}

\section{The Edwards model}

According to the Edwards model \cite{Edw65},
the Hamiltonian $H$ of a self-avoiding
chain equals the sum of the Hamiltonian of a standard Gaussian
chain $H_{0}$ and a quantity proportional to the number
of self-intersections per unit length $\nu^{\ast}$.
Equations (\ref{1}), (\ref{4}) and (\ref{11}) imply that
\begin{eqnarray}\label{15}
\hspace*{-4 mm}H({\bf r})&=& k_{B}T \biggl [ \frac{d}{2b_{0}}\int_{0}^{L}
\Bigl ( \frac{{\rm d} {\bf r}}{{\rm d} t}(t)\Bigr )^{2} {\rm d} t
+ \frac{v b_{0}^{d-1}}{2 L}
\nonumber\\
\hspace*{-4 mm}&&\times
\int_{0}^{L} {\rm d} t\int_{0}^{L}
\rho (t-\tau) \delta({\bf r}(t)-{\bf r}(\tau)){\rm d} \tau \biggr ],
\end{eqnarray}
where $v$ stands for the strength of self-repellent interactions.
The scattering function is given by
\begin{equation}\label{16}
S({\bf q})=\frac{1}{L^{2}}\int_{0}^{L} {\rm d}t \int_{0}^{L}
\Bigl \langle
\exp \Bigl (\imath {\bf q}\cdot ({\bf r}(t)-{\bf r}(\tau) \Bigr )
\Bigr \rangle {\rm d}\tau,
\end{equation}
where ${\bf q}$ is the scattering vector.
For a chain with the Hamiltonian $H({\bf r})$
and the Green function $G({\bf Q})$, the average value
$\langle\Phi\rangle$ of a functional $\Phi$ reads
\begin{eqnarray}\label{17}
\langle \Phi\rangle &=& \int \langle\Phi\rangle_{\bf Q}G({\bf Q}) {\rm
d}{\bf Q},
\nonumber\\
\langle \Phi\rangle_{\bf Q} &=& \frac{1}{G({\bf Q})}
\int_{{\bf r}(0)={\bf 0}}^{{\bf r}(L)={\bf Q}}
\Phi ({\bf r}(t))
\nonumber\\
&&\times \exp
\Bigl (-\frac{H({\bf r}(t))}{k_{B}T}\Bigr )
{\cal D}[{\bf r}(t)],
\end{eqnarray}
where ${\cal D}$ is a measure on the set of vector-functions
${\bf r}(t)$ obeying the boundary conditions in the
path integral \cite{Kle95}.
Combination of Eqs. (\ref{16}) and (\ref{17}) implies that
\begin{equation}\label{18}
S({\bf q})=\int \Sigma({\bf q},{\bf Q}) {\rm d} {\bf Q},
\end{equation}
where
\begin{eqnarray}\label{19}
\hspace*{-5 mm}\Sigma({\bf q},{\bf Q})&=& \frac{1}{L^{2}}
\int_{{\bf r}(0)={\bf 0}}^{{\bf r}(L)={\bf Q}}
\exp \Bigl ( -\frac{H({\bf r}(t))}{k_{B}T}\Bigr )
{\cal D}[{\bf r}(t)]
\nonumber\\
\hspace*{-5 mm}&& \times
\biggl [ \int_{0}^{L} {\rm d}t \int_{0}^{L}
\exp \Bigl (\imath {\bf q}\cdot \Bigl ({\bf r}(t)-{\bf r}(\tau)
\Bigr )\Bigr )\biggr ].
\end{eqnarray}
According to Eq. (\ref{16}), the scattering function $S({\bf q})$
satisfies the normalization condition
\begin{equation}\label{20}
S({\bf 0})=1.
\end{equation}
The radius of gyration of a chain
\begin{equation}\label{22}
R_{\rm g}^{2} =\frac{1}{2L^{2}}\int_{0}^{L} {\rm d}t \int_{0}^{L}
\Bigl \langle \Bigl ({\bf r}(t)-{\bf r}(\tau)\Bigr )^{2}
\Bigr \rangle {\rm d}\tau
\end{equation}
is connected with the function $S({\bf q})$ by the formula
\begin{equation}\label{21}
R_{\rm g}^{2}=-\frac{1}{2} \Delta S({\bf 0}),
\end{equation}
where $\Delta$ stands for the Laplace operator.
For an isotropic scattering function that depends on $q^{2}$
only, $S({\bf q})={\cal S}(q^{2})$,
Eq. (\ref{21}) is substantially simplified,
\begin{equation}\label{23}
R_{\rm g}^{2}=-d {\cal S}^{\prime}(0),
\end{equation}
where the prime stands for the derivative.

\section{Expansion of the scattering function}

Expanding the exponent of the Hamiltonian $H$ into the
Taylor series in $v$, we find from Eqs. (\ref{1}) and (\ref{15}) that
\begin{eqnarray}\label{24}
&&\exp \Bigl (-\frac{H({\bf r})}{k_{B}T}\Bigr )
=\exp \Bigl (-\frac{H_{0}({\bf r})}{k_{B}T}\Bigr )
\sum_{n=0}^{\infty} \frac{(-1)^{n}}{n!}
\Bigl (\frac{vb_{0}^{d-1}}{2L }\Bigr )^{n}
\nonumber\\
&&\times
\biggl [ \int_{0}^{L} {\rm d}t\int_{0}^{L}
\rho(t-\tau)
\delta \Bigl ({\bf r}(t)-{\bf r}(\tau)\Bigr )
{\rm d}\tau \biggr ]^{n}.
\end{eqnarray}
Inserting expression (\ref{24}) into Eq. (\ref{19}),
changing the order of integration and summation,
and using Eq. (\ref{17}), we obtain
\begin{eqnarray}\label{25}
&&\Sigma({\bf q},{\bf Q}) = \frac{G_{0}({\bf Q})}{L^{2}}
\sum_{n=0}^{\infty} \frac{(-1)^{n}}{n!}
\Bigl (\frac{vb_{0}^{d-1}}{2L }\Bigr )^{n}
\nonumber\\
&&\times
\biggl \langle
\Bigl [ \int_{0}^{L} {\rm d}t \int_{0}^{L} \exp \Bigl
(\imath {\bf q}\cdot \Bigl ({\bf r}(t)-{\bf r}(\tau)\Bigr )
\Bigr )\Bigr ]
\nonumber\\
&&\times
\Bigl [ \int_{0}^{L} {\rm d}t\int_{0}^{L}
\rho(t-\tau)
\delta \Bigl ({\bf r}(t)-{\bf r}(\tau)\Bigr )
{\rm d}\tau \Bigr ]^{n}\biggr \rangle_{\bf Q}^{0},
\end{eqnarray}
where $\langle\cdot\rangle_{\bf Q}^{0}$ stands for
averaging over configurations of a chain with
Hamiltonian (\ref{1}).
To derive an explicit expression for the $n$th moment
$(n=0,1,\ldots)$
\begin{eqnarray}\label{26}
M_{n}&=& \biggl \langle \Bigl [ \int_{0}^{L} {\rm d}t
\int_{0}^{L} \exp \Bigl (\imath {\bf q}\cdot
\Bigl ({\bf r}(t)-{\bf r}(\tau)\Bigr )\Bigr )\Bigr ]
\nonumber\\
&&\times
\Bigl [ \int_{0}^{L} {\rm d}t\int_{0}^{L}
\delta \Bigl ({\bf r}(t)-{\bf r}(\tau)\Bigr )
{\rm d}\tau \Bigr ]^{n}\biggr \rangle_{\bf Q}^{0},
\end{eqnarray}
we use the identity
\begin{eqnarray*}
&& \Bigl [ \int_{0}^{L} {\rm d}t\int_{0}^{L}
\rho(t-\tau)
\delta \Bigl ({\bf r}(t)-{\bf r}(\tau)\Bigr )
{\rm d}\tau \Bigr ]^{n}
\nonumber\\
&&=\int_{0}^{L} {\rm d}t_{1} \int_{0}^{L} {\rm d}\tau_{1}
\ldots
\int_{0}^{L} {\rm d}t_{n} \int_{0}^{L} {\rm d}\tau_{n}
\nonumber\\
&&\times
\prod_{m=1}^{n} \rho(t_{m}-\tau_{m})
\delta \Bigl ({\bf r}(t_{m})-{\bf r}(\tau_{m})\Bigr ),
\end{eqnarray*}
insert Eq. (\ref{3}), and obtain
\begin{eqnarray*}
&& M_{n}=\frac{1}{(2\pi)^{nd}}\prod_{m=1}^{n} \int {\rm d} {\bf k}_{m}
\int_{0}^{L} {\rm d}t_{1} \int_{0}^{L} {\rm d}\tau_{1}
\ldots
\nonumber\\
&&\times
\int_{0}^{L} {\rm d}t_{n} \int_{0}^{L} {\rm d}\tau_{n}
\int_{0}^{L} {\rm d}t_{n+1} \int_{0}^{L} {\rm d}\tau_{n+1}
\prod_{m=1}^{n} \rho(t_{m}-\tau_{m})
\nonumber\\
&&\times
\Bigl \langle
\exp \Bigl [ -\imath \sum_{m=1}^{n+1} {\bf k}_{m} \cdot
\Bigl ({\bf r}(t_{m})-{\bf r}(\tau_{m})\Bigr )\Bigr ]
\Bigr \rangle_{\bf Q}^{0},
\end{eqnarray*}
where
\begin{equation}\label{27}
t_{n+1}=t,
\quad
\tau_{n+1}=\tau,
\quad
{\bf k}_{n+1}=-{\bf q}.
\end{equation}
\begin{widetext}
Ordering the set of points $\{ t_{m},\tau_{m}\}$
as $t_{1}>\tau_{1}>t_{2}>\tau_{2}\ldots >t_{n+1}>\tau_{n+1}$
and bearing in mind that for any integrable function
$S(t_{1},\tau_{1},\ldots,t_{n+1},\tau_{n+1})$,
\begin{eqnarray*}
\int_{0}^{L} {\rm d}t_{1}\int_{0}^{L} {\rm d}\tau_{1}
\ldots
\int_{0}^{L} {\rm d} t_{n+1}\int_{0}^{L}
S {\rm d}\tau_{n+1}=(2n+2)!
\int_{0}^{L} {\rm d}t_{1} \int_{0}^{t_{1}} {\rm d}\tau_{1}
\ldots
\int_{0}^{\tau_{n}} {\rm d}t_{n+1} \int_{0}^{t_{n+1}}
S{\rm d} \tau_{n+1},
\end{eqnarray*}
we find that
\begin{eqnarray}\label{28}
M_{n} &=& \frac{(2n+2)!}{(2\pi)^{nd}}\prod_{m=1}^{n} \int {\rm d} {\bf k}_{m}
\int_{0}^{L} {\rm d}t_{1} \int_{0}^{t_{1}} {\rm d}\tau_{1}
\ldots
\int_{0}^{\tau_{n}} {\rm d}t_{n+1} \int_{0}^{t_{n+1}}
{\rm d}\tau_{n+1}
\prod_{m=1}^{n} \rho(t_{m}-\tau_{m})
\nonumber\\
&&\times
\Bigl \langle
\exp \Bigl [ -\imath \sum_{m=1}^{n+1} {\bf k}_{m} \cdot
\Bigl ({\bf r}(t_{m})-{\bf r}(\tau_{m})\Bigr )\Bigr ]
\Bigr \rangle_{\bf Q}^{0} .
\end{eqnarray}
For an ordered set of points $\{ t_{m},\tau_{m} \}$,
the average in Eq. (\ref{28}) reads
\begin{eqnarray}\label{29}
\Bigl \langle
\exp \Bigl [ -\imath \sum_{m=1}^{n+1} {\bf k}_{m} \cdot
\Bigl ({\bf r}(t_{m})-{\bf r}(\tau_{m})\Bigr )\Bigr ]
\Bigr \rangle_{\bf Q}^{0}
=\prod_{l=1}^{d} \exp \Bigl ( -\frac{b_{0}}{2d} \tilde{\bf k}_{l}
\cdot \tilde{\bf C}(\tilde{\bf T}) \cdot \tilde{\bf k}_{l}
-\frac{\imath}{L} Q_{l} \tilde{\bf k}_{l}\cdot \tilde{\bf T} \Bigr ),
\end{eqnarray}
where $Q_{l}$ is the $l$th component of the vector ${\bf Q}$
in a Cartesian coordinate frame,
$k_{ml}$ is the $l$th component of the vector ${\bf k}_{m}$,
and $T_{m}=t_{m}-\tau_{m}$ ($m=1,\ldots,n+1$).
The $(n+1)$-vectors $\tilde{\bf k}_{l}$ and $\tilde{\bf T}$
and the $(n+1)\times (n+1)$-matrix $\tilde{\bf C}({\bf T})$
are given by
\begin{eqnarray}\label{30}
\tilde{\bf k}_{l} &=& \left [
\begin{array}{c}
k_{1l}\\
k_{2l}\\
\ldots\\
k_{nl}\\
k_{n+1\; l}
\end{array}\right ]
=\left [
\begin{array}{c}
\bar{\bf k}_{l}\\
-q_{l}
\end{array}\right ],
\qquad
\tilde{\bf T}=\left [
\begin{array}{c}
T_{1}\\
T_{2}\\
\ldots\\
T_{n}\\
T_{n+1}
\end{array}\right ]
=\left [
\begin{array}{c}
\bar{\bf T}\\
t-\tau
\end{array}\right ],
\\
\label{31}
\tilde{\bf C} &=&
\left [\begin{array}{ccccc}
T_{1}(1-T_{1}/L) & -T_{1}T_{2}/L    & \ldots & -T_{1}T_{n}/L   & -T_{1}T_{n+1}/L\\
-T_{1}T_{2}/L    & T_{2}(1-T_{2}/L) & \ldots & -T_{2}T_{n}/L   & -T_{2}T_{n+1}/L\\
-T_{1}T_{3}/L    & -T_{2}T_{3}/L    & \ldots & -T_{3}T_{n}/L   & -T_{3}T_{n+1}/L\\
\ldots           & \ldots           & \ldots & \ldots          & \ldots \\
-T_{1}T_{n}/L    & -T_{2}T_{n}/L    & \ldots & T_{n}(1-T_{n}/L)& -T_{n}T_{n+1}/L\\
-T_{1}T_{n+1}/L  & -T_{2}T_{n+1}/L  & \ldots & -T_{n}T_{n+1}/L & T_{n+1}(1-T_{n+1}/L)
\end{array}\right ]
\nonumber\\
&=&
\left [\begin{array}{cc}
{\bf C} & -{\bf T}(t-\tau)/L\\
-{\bf T}(t-\tau)/L & (t-\tau)(1-(t-\tau)/L)
\end{array}\right ].
\end{eqnarray}
Taking into account that
\begin{eqnarray}\label{32}
\prod_{l=1}^{d} \exp \Bigl ( -\frac{b_{0}}{2d} \tilde{\bf k}_{l}
\cdot \tilde{\bf C} (\tilde{\bf T})\cdot \tilde{\bf k}_{l}
-\frac{\imath}{L} Q_{l} \tilde{\bf k}_{l}\cdot \tilde{\bf T} \Bigr )
&=&\exp \biggl [-\sum_{l=1}^{d}\biggl (
\frac{b_{0}}{2d} \bar{\bf k}_{l}
\cdot {\bf C}({\bf T}) \cdot \bar{\bf k}_{l}
+\frac{\imath}{L} \Bigl (Q_{l}-\imath q_{l}\frac{b_{0}}{d}
(t-\tau)\Bigr )\bar{\bf k}_{l}\cdot {\bf T} \biggr )
\biggr ]
\nonumber\\
&&\times
\exp \biggl [ -\frac{t-\tau}{L}
\biggl (\frac{b_{0}L q^{2}}{2d}\Bigl (1-\frac{t-\tau}{L}\Bigr )
-\imath {\bf Q}\cdot {\bf q} \biggr ) \biggr ].
\end{eqnarray}
we find from Eqs. (\ref{28}) and (\ref{29}) that
\begin{eqnarray*}
M_{n} &=& \frac{(2n+2)!}{(2\pi)^{nd}}
\prod_{l=1}^{d} \int {\rm d} \bar{\bf k}_{l}
\int_{0}^{L} {\rm d}t_{1} \int_{0}^{t_{1}} {\rm d}\tau_{1}
\ldots
\int_{0}^{\tau_{n}} {\rm d}t \int_{0}^{t} {\rm d}\tau
\exp \biggl [-\sum_{l=1}^{d}\biggl (
\frac{b_{0}}{2d} \bar{\bf k}_{l}
\cdot {\bf C}({\bf T}) \cdot \bar{\bf k}_{l}
\nonumber\\
&&
+\frac{\imath}{L} \Bigl (Q_{l}-\imath q_{l}\frac{b_{0}}{d}
(t-\tau)\Bigr )\bar{\bf k}_{l}\cdot {\bf T} \biggr )
\biggr ]
\prod_{m=1}^{n} \rho(t_{m}-\tau_{m})
\exp \biggl [ -\frac{t-\tau}{L}
\biggl (\frac{b_{0}L q^{2}}{2d}\Bigl (1-\frac{t-\tau}{L}\Bigr )
-\imath {\bf Q}\cdot {\bf q} \biggr ) \biggr ],
\end{eqnarray*}
\end{widetext}
where we replaced integration over ${\bf k}_{m}$ with
integration over $\bar{\bf k}_{l}$ and returned to the
initial notation $t$ and $\tau$.
Calculating the Gaussian integral over $\bar{\bf k}_{l}$
and bearing in mind that
$\int_{0}^{\tau_{n}}{\rm d}t\int_{0}^{t} \varphi(t-\tau)
{\rm d}\tau
=\int_{0}^{\tau_{n}} (\tau_{n}-z) \varphi(z)
{\rm d} z$
for any function $\varphi(t-\tau)$, we obtain
\begin{eqnarray}\label{33}
M_{n} &=&
\int_{0}^{L} {\rm d}t_{1} \int_{0}^{t_{1}} {\rm d}\tau_{1}
\ldots
\int_{0}^{\tau_{n-1}} {\rm d}t_{n} \int_{0}^{t_{n}} {\rm d}\tau_{n}
\nonumber\\
&&\times
\int_{0}^{\tau_{n}} (\tau_{n}-z) R_{n}({\bf T},z) {\rm d}z,
\end{eqnarray}
where
\begin{eqnarray}\label{34}
\hspace*{-7 mm}&& R_{n}({\bf T},z) =(2n+2)!
\Bigl (\frac{d}{2\pi b_{0}}\Bigr )^{\frac{nd }{2}}
\Bigl ( \frac{1}{\det {\bf C({\bf T})}}\Bigr )^{\frac{d}{2}}
\nonumber\\
\hspace*{-7 mm}&&\times
\exp \Bigl [-\frac{d}{2b_{0}L^{2}}
\Bigl ({\bf Q}-\frac{\imath b_{0}}{d} {\bf q}z\Bigr )^{2}
{\bf T}\cdot {\bf C}^{-1}({\bf T})\cdot {\bf T}\Bigr ]
\nonumber\\
\hspace*{-7 mm}&&\times
\prod_{m=1}^{n} \rho(T_{m})
\exp \Bigl [ -\frac{z}{L} \Bigl (\frac{b_{0}L q^{2}}{2d}
\Bigl (1-\frac{z}{L}\Bigr ) -\imath {\bf Q}\cdot {\bf q} \Bigr )
\Bigr ],
\end{eqnarray}
and we omit the arguments ${\bf q}$ and ${\bf Q}$ of
the function $R_{n}$.
At $n=0$, Eqs. (\ref{33}) and (\ref{34}) read
\begin{eqnarray}\label{35}
&&\biggl \langle \int_{0}^{L} {\rm d}t
\int_{0}^{L} \exp \Bigl (\imath {\bf q}\cdot
\Bigl ({\bf r}(t)-{\bf r}(\tau)\Bigr )\Bigr )
\biggr \rangle_{\bf Q}^{0}
\nonumber\\
&&= \int_{0}^{L} (L-z) R_{0}({\bf T},z) {\rm d}z
\end{eqnarray}
with
\begin{equation}\label{36}
R_{0}({\bf T},z) = 2
\exp \Bigl [ -\frac{z}{L} \Bigl (\frac{b_{0}L q^{2}}{2d}
\Bigl (1-\frac{z}{L}\Bigr ) -\imath {\bf Q}\cdot {\bf q} \Bigr )
\Bigr ].
\end{equation}
To reduce the number of integrations in Eq. (\ref{33}),
we infer by induction that for any $n=1,2,\ldots$,
\begin{eqnarray}\label{37}
&& \int_{0}^{L} {\rm d}t_{1}
\int_{0}^{t_{1}} {\rm d} \tau_{1}
\ldots \int_{0}^{\tau_{n-1}} {\rm d}t_{n}
\int_{0}^{t_{n}} {\rm d}\tau_{n}
\int_{0}^{\tau_{n}} (\tau_{n}-z)
\nonumber\\
&&\times
R_{n}({\bf T},z) {\rm d}z
= \frac{1}{(n+1)!} \int_{0}^{L} {\rm d}T_{1}
\int_{0}^{L-T_{1}}
{\rm d}T_{2} \ldots
\nonumber\\
&&\times
\int_{0}^{L-\sum_{m=1}^{n-1}T_{m}} {\rm d}T_{n}
\int_{0}^{L-\sum_{m=1}^{n}T_{m}}
\Bigl (L-\sum_{m=1}^{n}T_{m}
\nonumber\\
&& -z\Bigr )^{n+1}R_{n}({\bf T},z) {\rm d}z .
\end{eqnarray}
Introducing the new variables
\begin{equation}\label{38}
u_{m}=L-\sum_{k=1}^{m} T_{k}
\qquad
(m=1,2,\ldots,n),
\end{equation}
we present Eq. (\ref{37}) in the form
\begin{eqnarray}\label{39}
&& \int_{0}^{L} {\rm d}t_{1}
\int_{0}^{t_{1}} {\rm d} \tau_{1}
\ldots \int_{0}^{\tau_{n-1}} {\rm d}t_{n}
\int_{0}^{t_{n}} {\rm d}\tau_{n}
\int_{0}^{\tau_{n}} (\tau_{n}-z)
\nonumber\\
&&\times
R_{n}({\bf T},z) {\rm d}z
=\frac{1}{(n+1)!} \int_{0}^{L} {\rm d} u_{1}
\int_{0}^{u_{1}} {\rm d} u_{2} \ldots
\nonumber\\
&&\times
\int_{0}^{u_{n-1}} {\rm d} u_{n}
\int_{0}^{u_{n}} (u_{n}-z)^{n+1} R_{n}({\bf u},z) {\rm d}z,
\end{eqnarray}
where $R_{n}({\bf u},z)$ is obtained from $R_{n}({\bf T},z)$ after
substitution of the expressions
\begin{equation}\label{40}
T_{1}=L-u_{1},
\quad
T_{m}=u_{m-1}-u_{m}
\quad
(m=2,\ldots,n),
\end{equation}
corresponding to the inverse transformation.
To find an explicit expression for the function $R_{n}({\bf u},z)$,
we use the formulas
\begin{eqnarray}\label{41}
\det {\bf C}({\bf T}) &=& \Bigl (1-\frac{1}{L}
\sum_{m=1}^{n} T_{m}\Bigr )
\prod_{m=1}^{n} T_{m}
\nonumber\\
&=& \frac{1}{L}(L-u_{1})u_{n}
\prod_{m=2}^{n} (u_{m-1}-u_{m}),
\nonumber\\
{\bf T}\cdot {\bf C}^{-1}({\bf T})\cdot {\bf T}
&=& L\sum_{m=1}^{n} T_{m} \Bigl (L-\sum_{m=1}^{n} T_{m}
\Bigr )^{-1}
\nonumber\\
&=& L\frac{L-u_{n}}{u_{n}},
\end{eqnarray}
that follow from Eqs. (\ref{30}), (\ref{31}) and (\ref{40}).
Insertion of expressions (\ref{41}) into Eq. (\ref{34})
results in
\begin{eqnarray}\label{42}
&& R_{n}({\bf T},z) =(2n+2)!
\Bigl (\frac{d}{2\pi b_{0}} \Bigr )^{\frac{dn}{2}}
L^{\frac{d}{2}}
\nonumber\\
&&\times
\frac{\rho(L-u_{1})\rho(u_{1}-u_{2})\ldots \rho(u_{n-1}-u_{n})}
{(L-u_{1})^{\frac{d}{2}} (u_{1}-u_{2})^{\frac{d}{2}} \ldots
(u_{n-1}-u_{n})^{\frac{d}{2}}u_{n}^{\frac{d}{2}}}
\nonumber\\
&& \times
\exp \Bigl [-\frac{d(L-u_{n})}{2b_{0}L u_{n}}
\Bigl ({\bf Q}-\frac{\imath b_{0}}{d} {\bf q}z\Bigr )^{2}
\nonumber\\
&& -\frac{z}{L} \Bigl (\frac{b_{0}L q^{2}}{2d}
\Bigl (1-\frac{z}{L}\Bigr ) -\imath {\bf Q}\cdot {\bf q} \Bigr )
\Bigr ].
\end{eqnarray}
It follows from Eqs. (\ref{25}), (\ref{26}), (\ref{33}),
(\ref{35}), (\ref{36}), (\ref{39}) and (\ref{42}) that
\begin{widetext}
\begin{eqnarray}\label{43}
\Sigma({\bf q},{\bf Q}) &=&
\frac{G_{0}({\bf Q})}{L^{2}}
\biggl \{ 2 \int_{0}^{L} (L-z)
\exp \Bigl [ -\frac{z}{L} \Bigl (\frac{b_{0}L q^{2}}{2d}
\Bigl (1-\frac{z}{L}\Bigr ) -\imath {\bf Q}\cdot {\bf q} \Bigr )
\Bigr ] {\rm d}z
+ L^{\frac{d}{2}}
\sum_{n=1}^{\infty} \frac{(-1)^{n}}{n!}
\Bigl (\frac{vb_{0}^{d-1}}{2L }\Bigr )^{n}
\frac{(2n+2)!}{(n+1)!}
\nonumber\\
&&\times
\Bigl (\frac{d}{2\pi b_{0}} \Bigr )^{\frac{dn}{2}}
\int_{0}^{L} \frac{\rho(L-u_{1}){\rm d} u_{1}}{(L-u_{1})^{\frac{d}{2}}}
\int_{0}^{u_{1}} \frac{\rho(u_{1}-u_{2}) {\rm d} u_{2}}
{(u_{1}-u_{2})^{\frac{d}{2}}}
\ldots \int_{0}^{u_{n-1}} \frac{\rho(u_{n-1}-u_{n}){\rm d} u_{n}}
{(u_{n-1}-u_{n})^{\frac{d}{2}}u_{n}^{\frac{d}{2}}}
\int_{0}^{u_{n}} (u_{n}-z)^{n+1}
\nonumber\\
&&\times
\exp \Bigl [-\frac{d(L-u_{n})}{2b_{0}L u_{n}}
\Bigl ({\bf Q}-\frac{\imath b_{0}}{d} {\bf q}z\Bigr )^{2}
-\frac{z}{L} \Bigl (\frac{b_{0}L q^{2}}{2d}
\Bigl (1-\frac{z}{L}\Bigr ) -\imath {\bf Q}\cdot {\bf q} \Bigr )
\Bigr ] {\rm d}z \biggr \}.
\end{eqnarray}
\end{widetext}

\section{The Laplace transform}

The Green function for a Gaussian chain in a $d$-dimensional
space reads
\begin{equation}\label{44}
G_{0}({\bf Q})=\bar{G} \Bigl (\frac{d}{2\pi b_{0}L}\Bigr )^{\frac{d}{2}}
\exp \Bigl (-\frac{d Q^{2}}{2 b_{0}L}\Bigr ).
\end{equation}
Equation (\ref{44}) differs from Eq. (\ref{9}) by
the pre-factor $\bar{G}$.
This coefficient will be found later from
normalization condition (\ref{20}).
Substitution of expressions (\ref{43}) and (\ref{44})
into Eq. (\ref{18}) results in
\begin{widetext}
\begin{eqnarray*}
S({\bf q}) &=& \frac{\bar{G}}{L^{2}}
\Bigl (\frac{d}{2\pi b_{0}}\Bigr )^{\frac{d}{2}}
\biggl \{ \frac{2}{L^{\frac{d}{2}}} \int_{0}^{L} (L-z)
\exp \Bigl ( -\frac{b_{0}q^{2}z(L-z)}{2dL} \Bigr ) {\rm d}z
\int \exp \Bigl (-\frac{d Q^{2}}{2 b_{0}L}
+\frac{\imath z}{L} {\bf Q}\cdot {\bf q} \Bigr )
{\rm d}{\bf Q}
+ \sum_{n=1}^{\infty} \frac{(-1)^{n}}{n!}
\Bigl (\frac{vb_{0}^{d-1}}{2 L}\Bigr )^{n}
\nonumber\\
&&\times
\frac{(2n+2)!}{(n+1)!}
\Bigl (\frac{d}{2\pi b_{0}} \Bigr )^{\frac{dn}{2}}
\int_{0}^{L} \frac{\rho(L-u_{1}){\rm d} u_{1}}{(L-u_{1})^{\frac{d}{2}}}
\int_{0}^{u_{1}} \frac{\rho(u_{1}-u_{2}) {\rm d} u_{2}}
{(u_{1}-u_{2})^{\frac{d}{2}}}
\ldots \int_{0}^{u_{n-1}} \frac{\rho(u_{n-1}-u_{n}){\rm d} u_{n}}
{(u_{n-1}-u_{n})^{\frac{d}{2}}u_{n}^{\frac{d}{2}}}
\int_{0}^{u_{n}} (u_{n}-z)^{n+1}
\nonumber\\
&&\times
\exp \Bigl (-\frac{b_{0}q^{2}z(u_{n}-z)}{2d u_{n}} \Bigr )
{\rm d}z
\int \exp \Bigl (-\frac{d Q^{2}}{2 b_{0}u_{n}}
+\frac{\imath z}{u_{n}} {\bf Q}\cdot {\bf q} \Bigr)
{\rm d}{\bf Q} \biggr \}.
\end{eqnarray*}
Calculation of the Gaussian integrals yields
\begin{eqnarray}\label{5-45}
\frac{S({\bf q})}{\bar{G}} &=&
S_{D}\Bigl (\frac{b_{0}L}{2d}q^{2}\Bigr )
+\frac{1}{L^{2}}
\sum_{n=1}^{\infty} \frac{(-1)^{n}}{n!}
\Bigl (\frac{vb_{0}^{d-1}}{2 L}\Bigr )^{n}
\frac{(2n+2)!}{(n+1)!}
\Bigl (\frac{d}{2\pi b_{0}} \Bigr )^{\frac{dn}{2}}
\int_{0}^{L} \frac{\rho(L-u_{1}){\rm d} u_{1}}{(L-u_{1})^{\frac{d}{2}}}
\int_{0}^{u_{1}} \frac{\rho(u_{1}-u_{2}) {\rm d} u_{2}}
{(u_{1}-u_{2})^{\frac{d}{2}}}
\ldots
\nonumber\\
&& \times
\int_{0}^{u_{n-1}} \frac{\rho(u_{n-1}-u_{n}){\rm d} u_{n}}
{(u_{n-1}-u_{n})^{\frac{d}{2}}}
\int_{0}^{u_{n}} (u_{n}-z)^{n+1}
\exp \Bigl (-\frac{b_{0}z q^{2}}{2d} \Bigr ) {\rm d}z ,
\end{eqnarray}
\end{widetext}
where
\begin{eqnarray}\label{46}
S_{D}\Bigl (\frac{b_{0}L}{2d}q^{2}\Bigr )
&=& \frac{2}{L^{2}} \int_{0}^{L} \exp
\Bigl (-\frac{b_{0}q^{2}}{2d}z \Bigr ) (L-z) {\rm d}z
\nonumber\\
&=& 2 \int_{0}^{1} \exp \Bigl (-\frac{b_{0}Lq^{2}}{2d}(1-x)\Bigr )
x {\rm d}x
\end{eqnarray}
is the Debye function.
Introducing the notation
\begin{equation}\label{47}
P(L,{\bf q})=L^{2}\biggl [
\frac{S({\bf q})}{\bar{G}}
-S_{D}\Bigl (\frac{b_{0}L}{2d}q^{2}\Bigr )\biggr ],
\end{equation}
we infer that
\begin{eqnarray}\label{48}
P(L,{\bf q}) &=& \sum_{n=1}^{\infty} \frac{(-1)^{n}}{n!}
\Bigl (\frac{vb_{0}^{d-1}}{2 L}\Bigr )^{n}
\frac{(2n+2)!}{(n+1)!}
\Bigl (\frac{d}{2\pi b_{0}} \Bigr )^{\frac{dn}{2}}
\nonumber\\
&&\times
\int_{0}^{L} \frac{\rho(L-u_{1}){\rm d} u_{1}}{(L-u_{1})^{\frac{d}{2}}}
\int_{0}^{u_{1}} \frac{\rho(u_{1}-u_{2}) {\rm d} u_{2}}
{(u_{1}-u_{2})^{\frac{d}{2}}}
\ldots
\nonumber\\
&&\times
\int_{0}^{u_{n-1}} \frac{\rho(u_{n-1}-u_{n}){\rm d} u_{n}}
{(u_{n-1}-u_{n})^{\frac{d}{2}}}
\int_{0}^{u_{n}} (u_{n}-z)^{n+1}
\nonumber\\
&&\times
\exp \Bigl (-\frac{b_{0}z q^{2}}{2d} \Bigr ) {\rm d}z.
\end{eqnarray}
To derive an explicit formula for the function $P(L,{\bf q})$,
the following technique is employed:
(i) first, we perform the Laplace transform of Eq. (\ref{48}),
(ii) afterwards, we sum the infinite series
in the transformed space, and
(iii) finally, we perform the inverse Laplace transform.

Denote by $\hat{P}(\lambda,{\bf q})$ the Laplace transform
of the function $P(L,{\bf q})$ with respect to $L$.
Multiplying Eq. (\ref{48}) by $\exp(-\lambda L)$ and integrating
over $L$ from zero to infinity, we find that
\begin{eqnarray}\label{49}
&&\hat{P}(\lambda,{\bf q})
= \sum_{n=1}^{\infty} \frac{(-1)^{n}}{n!}
\Bigl (\frac{vb_{0}^{d-1}}{2 L}\Bigr )^{n}
\frac{(2n+2)!}{(n+1)!}
\Bigl (\frac{d}{2\pi b_{0}} \Bigr )^{\frac{dn}{2}}
\nonumber\\
&&\times
\int_{0}^{\infty} \exp (-\lambda L) {\rm d}L
\int_{0}^{L} \frac{\rho(L-u_{1}){\rm d} u_{1}}{(L-u_{1})^{\frac{d}{2}}}
\nonumber\\
&&\times
\int_{0}^{u_{1}} \frac{\rho(u_{1}-u_{2}) {\rm d} u_{2}}
{(u_{1}-u_{2})^{\frac{d}{2}}}
\ldots \int_{0}^{u_{n-1}} \frac{\rho(u_{n-1}-u_{n}){\rm d} u_{n}}
{(u_{n-1}-u_{n})^{\frac{d}{2}}}
\nonumber\\
&&\times
\int_{0}^{u_{n}} (u_{n}-z)^{n+1}
\exp \Bigl (-\frac{b_{0}z q^{2}}{2d} \Bigr ) {\rm d}z.
\end{eqnarray}
According to the convolution theorem for the Laplace
transform, for any function $\varphi (u)$,
\begin{eqnarray}\label{50}
&& \int_{0}^{\infty} \exp (-\lambda L) {\rm d}L
\int_{0}^{L} \frac{\rho(L-u_{1})}{(L-u_{1})^{\frac{d}{2}}}
\varphi(u_{1}){\rm d} u_{1}
\nonumber\\
&& =\Lambda(\lambda)\int_{0}^{\infty} \varphi(u_{1})
\exp (-\lambda u_{1}) {\rm d}u_{1},
\end{eqnarray}
where
\begin{equation}\label{51}
\Lambda(\lambda)=\int_{0}^{\infty} \rho(x) x^{-\frac{d}{2}}
\exp (-\lambda x) {\rm d}x.
\end{equation}
Applying Eq. (\ref{50}) $n$ times, we obtain
\begin{eqnarray}\label{52}
&& \int_{0}^{\infty} \exp (-\lambda L) {\rm d}L
\int_{0}^{L} \frac{\rho(L-u_{1}){\rm d} u_{1}}{(L-u_{1})^{\frac{d}{2}}}
\nonumber\\
&&\times
\int_{0}^{u_{1}} \frac{\rho(u_{1}-u_{2}) {\rm d} u_{2}}
{(u_{1}-u_{2})^{\frac{d}{2}}}
\ldots \int_{0}^{u_{n-1}} \frac{\rho(u_{n-1}-u_{n}){\rm d} u_{n}}
{(u_{n-1}-u_{n})^{\frac{d}{2}}}
\nonumber\\
&&\times
\int_{0}^{u_{n}} (u_{n}-z)^{n+1}
\exp \Bigl (-\frac{b_{0}z q^{2}}{2d} \Bigr ) {\rm d}z
\nonumber\\
&&=(n+1)!\frac{\Lambda^{n}(\lambda)}{\lambda^{n+2}}
\Bigl (\lambda+\frac{b_{0}q^{2}}{2d} \Bigr )^{-1}.
\end{eqnarray}
Combination of Eqs. (\ref{49}) and (\ref{52}) implies that
\begin{eqnarray}\label{53}
\hat{P}(\lambda,{\bf q}) &=& \lambda^{-2}
\Bigl (\lambda+\frac{b_{0}q^{2}}{2d} \Bigr )^{-1}
\nonumber\\
&&\times
\sum_{n=1}^{\infty} (-1)^{n}
\frac{(2n+2)!}{n!}
\Bigl (\frac{v_{0}\Lambda(\lambda)}{2\lambda}\Bigr )^{n}
\end{eqnarray}
with
\begin{equation}\label{54}
v_{0}=\frac{v}{b_{0} L}
\Bigl (\frac{d b_{0}}{2\pi} \Bigr )^{\frac{d}{2}}.
\end{equation}
The pre-factor in Eq. (\ref{53}) reads
\[
a_{n}=\frac{(2n+2)!}{n!}
=2(n+1)\frac{\Gamma(2(n+1))}{\Gamma(n+1)}.
\]
Application of the Legendre duplication formula for the
gamma-function yields
\[
a_{n}= \frac{4^{n+1}}{\sqrt{\pi}}\Bigl [
\Gamma \Bigl (n+\frac{5}{2}\Bigr )
-\frac{1}{2}\Gamma \Bigl (n+\frac{3}{2}\Bigr )\Bigr ].
\]
Insertion of this expression into Eq. (\ref{53}) results in
\begin{eqnarray*}
&& \hat{P}(\lambda,{\bf q}) = \frac{4}{\sqrt{\pi}\lambda^{2}}
\Bigl (\lambda+\frac{b_{0}q^{2}}{2d} \Bigr )^{-1}
\biggl [ \sum_{n=1}^{\infty} (-1)^{n} \Gamma\Bigl (n+\frac{5}{2}\Bigr )
\nonumber\\
&&\times (2V)^{n}
-\frac{1}{2}\sum_{n=1}^{\infty} (-1)^{n}
\Gamma\Bigl (n+\frac{3}{2}\Bigr ) (2V)^{n} \biggr ],
\end{eqnarray*}
where we use the shortcut notation
\begin{equation}\label{55}
V=\frac{v_{0}\Lambda(\lambda)}{\lambda }.
\end{equation}
Bearing in mind that for any $X$,
\begin{eqnarray*}
&& -\frac{1}{2} \sum_{n=1}^{\infty} (-1)^{n}
\Gamma\Bigl (n+\frac{3}{2}\Bigr )X^{n}
\nonumber\\
&&
=\frac{X}{2} \Bigl [\frac{3\sqrt{\pi}}{4}
+\sum_{n=1}^{\infty} (-1)^{n}
\Gamma\Bigl (n+\frac{5}{2}\Bigr )X^{n}\Bigr ],
\end{eqnarray*}
we find that
\begin{eqnarray}\label{56}
\hspace*{-5 mm}\hat{P}(\lambda,{\bf q}) &=& \frac{4}{\sqrt{\pi}\lambda^{2}}
\Bigl (\lambda+\frac{b_{0}q^{2}}{2d} \Bigr )^{-1}
\biggl [ \frac{3\sqrt{\pi}}{4}V
\nonumber\\
\hspace*{-5 mm}&& +(1+V)
\sum_{n=1}^{\infty} (-1)^{n} \Gamma\Bigl (n+\frac{5}{2}\Bigr )
(2V)^{n} \biggr ],
\end{eqnarray}
With reference to the generalized method of Borel's summation
\cite{SW94}, an infinite series
\begin{equation}\label{57}
\sum_{n=1}^{\infty} a_{n}
\end{equation}
is associated with the function
\begin{equation}\label{58}
\sigma(z)=\exp (-z) \sum_{n=1}^{\infty}
\frac{a_{n} z^{n+\frac{3}{2}}}{\Gamma(n+\frac{5}{2})}.
\end{equation}
Series (\ref{57}) is Borel summable, and its sum equal $\sigma_{0}$,
provided that the integral $\int_{0}^{s} \sigma(z) {\rm d}z$
exists for any $s\geq 0$ and converges to $\sigma_{0}$
as $s\to \infty$ in the classical sense.
The summation method with function (\ref{58}) is regular:
for any convergent series (\ref{57}),
$\sigma_{0} = \sum_{n=1}^{\infty} a_{n}$.
Applying the above technique to the infinite series in
Eq. (\ref{56}), we find that for any $X$,
\[
\sum_{n=1}^{\infty} (-1)^{n} \Gamma\Bigl (n+\frac{5}{2}\Bigr )X^{n}
= -\int_{0}^{\infty} \exp (-z) \frac{X z^{\frac{5}{2}}}{1+X z}
{\rm d}z,
\]
where we performed summation of the geometrical series.
The expression in square brackets in Eq. (\ref{56}) reads
\begin{eqnarray}\label{59}
&& V\biggl [ \frac{3\sqrt{\pi}}{4}-2(1+V)\int_{0}^{\infty}
\frac{\exp(-z) z^{\frac{5}{2}}}{1+2 Vz} {\rm d}z \biggr ]
\nonumber\\
&& =V \int_{0}^{\infty} \frac{1-2z}{1+2 Vz}
\exp (-z) z^{\frac{3}{2}} {\rm d}z ,
\end{eqnarray}
where we used the identity
\[
\frac{3\sqrt{\pi}}{4} =\int_{0}^{\infty} \exp(-z)
z^{\frac{3}{2}} {\rm d}z.
\]
Inserting expression (\ref{59}) into Eq. (\ref{56}) and using
Eq. (\ref{55}), we arrive at the formula
\begin{eqnarray}\label{60}
\hat{P}(\lambda,{\bf q}) &=& \frac{4v_{0}}{\sqrt{\pi}\lambda^{2}}
\Bigl (\lambda+\frac{b_{0}q^{2}}{2d} \Bigr )^{-1}
\nonumber\\
&&\times
\int_{0}^{\infty}
\frac{\Lambda(\lambda)(1-2z)}{\lambda +2v_{0}\Lambda(\lambda)z }
\exp(-z) z^{\frac{3}{2}}{\rm d}z
\end{eqnarray}
for the Laplace transform of the function $P(L,{\bf q})$.
Our aim now is to find the function
\[
P(L,{\bf q})={\cal L}^{-1}\Bigl (\hat{P}(\lambda,{\bf q})
\Bigr )_{\lambda\to L},
\]
where ${\cal L}^{-1}$ stands for the inverse Laplace transform.
Taking into account that
\[
{\cal L}^{-1}\biggl (
\Bigl (\lambda+\frac{b_{0}q^{2}}{2d} \Bigr )^{-1}
\biggr )_{\lambda\to l}
=\exp \Bigl (-\frac{b_{0}lq^{2}}{2d}\Bigr ),
\]
and applying the convolution theorem for the inverse
Laplace transform, we obtain
\begin{eqnarray}\label{61}
\hspace*{-5 mm}&& P(L,{\bf q}) =  \frac{4v_{0}}{\sqrt{\pi}}
\int_{0}^{\infty} \exp(-z)(1-2z) z^{\frac{3}{2}} {\rm d}z
\nonumber\\
\hspace*{-5 mm}&&\times
\int_{0}^{L} \exp \Bigl (-\frac{b_{0}(L-l)q^{2}}{2d}\Bigr )
{\cal L}^{-1} \Bigl (W(\lambda,z)\Bigr )_{\lambda\to l} {\rm d}l,
\end{eqnarray}
where
\begin{equation}\label{62}
W(\lambda,z)=\frac{\Lambda(\lambda)\lambda^{-2}}
{\lambda+2v_{0}z \Lambda(\lambda)}.
\end{equation}
As the inverse Laplace transform in Eq. (\ref{61}) cannot
be determined in the general form for an arbitrary integer $d$,
we consider particular cases $d=1$ and $d=3$.

\section{A self-avoiding chain in an one-dimensional space}

As the integral in Eq. (\ref{51}) with $d=1$ converges
without regularization, we set $\rho=1$ and find that
\begin{equation}\label{63}
\Lambda(\lambda)=\int_{0}^{\infty} x^{-\frac{1}{2}}
\exp (-\lambda x) {\rm d}x
=\sqrt{\frac{\pi}{\lambda}} .
\end{equation}
Insertion of expression (\ref{63}) into Eq. (\ref{62}) results in
\begin{eqnarray}\label{64}
W(\lambda,z) &=&\sqrt{\pi}\frac{\lambda^{-2}}
{\lambda^{\frac{3}{2}}+\phi(z)},
\nonumber\\
\phi(z) &=& 2v_{0}\sqrt{\pi}z
= v z \sqrt{\frac{2}{b_{0} L^{2}}}.
\end{eqnarray}
Taking into account that for any positive $\alpha$, $\beta$,
and $\phi$,
\begin{equation}\label{65}
{\cal L}^{-1}\biggl (
\frac{s^{\alpha-\beta}}{s^{\alpha}+\phi}
\biggr )_{s\to x}
=x^{\beta-1}E_{\alpha,\beta}(- \phi x^{\alpha}),
\end{equation}
where
\begin{equation}\label{66}
E_{\alpha,\beta}(x)=\sum_{n=0}^{\infty}
\frac{x^{n}}{\Gamma(\alpha n+\beta)}
\end{equation}
is the Mittag--Leffler function \cite{Erd55}, we infer that
\begin{equation}\label{67}
{\cal L}^{-1} \biggl ( W(\lambda,z)\biggr )_{\lambda\to l}
=\sqrt{\pi} l^{\frac{5}{2}} E_{\frac{3}{2},\frac{7}{2}}
\Bigl (-\phi(z) l^{\frac{3}{2}}\Bigr ).
\end{equation}
Combination of Eqs. (\ref{61}) and (\ref{67}) yields
\begin{eqnarray*}
&& P(L,{\bf q}) =  4v_{0}
\int_{0}^{\infty} \exp(-z)(1-2z) z^{\frac{3}{2}} {\rm d}z
\nonumber\\
&&\times
\int_{0}^{L} \exp \Bigl (-\frac{b_{0}(L-l)q^{2}}{2}\Bigr )
E_{\frac{3}{2},\frac{7}{2}}\Bigl (-\phi(z) l^{\frac{3}{2}}\Bigr )
l^{\frac{5}{2}} {\rm d}l.
\end{eqnarray*}
Setting $x=l/L$ and substituting expressions (\ref{54})
and (\ref{64}) into this equality, we arrive at the formula
\begin{eqnarray}\label{68}
\hspace*{-5 mm}&& P(L,{\bf q}) =  \frac{2 \chi L^{2}}{\sqrt{\pi}}
\int_{0}^{\infty} \exp(-z)(1-2z) z^{\frac{3}{2}} {\rm d}z
\nonumber\\
\hspace*{-5 mm}&&\times
\int_{0}^{1} \exp \Bigl (-\frac{b^{2}q^{2} }{2}(1-x)\Bigr )
E_{\frac{3}{2},\frac{7}{2}}(-\chi x^{\frac{3}{2}} z)
x^{\frac{5}{2}} {\rm d}x,
\end{eqnarray}
where
\begin{equation}\label{69}
b^{2}=b_{0}L,
\qquad
\chi =v \sqrt{\frac{2L}{b_{0}}}.
\end{equation}
Insertion of Eq. (\ref{68}) into Eq. (\ref{47}) yields
\begin{eqnarray}\label{70}
\hspace*{-4 mm}&& S({\bf q}) = \bar{G} \biggl [
S_{D}\Bigl (\frac{b^{2}q^{2}}{2}\Bigr )
+\frac{2 \chi}{\sqrt{\pi}}
\int_{0}^{1} \exp \Bigl (-\frac{b^{2}q^{2} }{2}(1-x)\Bigr )
\nonumber\\
\hspace*{-4 mm}&&\times
x^{\frac{5}{2}} {\rm d}x
\int_{0}^{\infty} \exp(-z)
E_{\frac{3}{2},\frac{7}{2}}(-\chi x^{\frac{3}{2}} z)
(1-2z) z^{\frac{3}{2}} {\rm d}z\biggr ] .
\end{eqnarray}
It follows from Eq. (\ref{66}) that for any $c>0$,
\begin{eqnarray}\label{71}
&& \int_{0}^{\infty} \exp(-z)
E_{\frac{3}{2},\frac{7}{2}}(-c z)
(1-2z) z^{\frac{3}{2}} {\rm d}z
\nonumber\\
&&=-2 \sum_{n=0}^{\infty}\frac{\Gamma (n+\frac{5}{2})(n+2)}
{\Gamma(\frac{3}{2}n+\frac{7}{2})} (-c)^{n}.
\end{eqnarray}
Substituting expression (\ref{46}) with $d=1$ into Eq. (\ref{70})
and using Eq. (\ref{71}), we obtain
\begin{eqnarray}\label{72}
S({\bf q}) &=& \frac{4 \bar{G}}{\sqrt{\pi}}
\int_{0}^{1} \exp \Bigl (-\frac{b^{2}q^{2}}{2}(1-x)\Bigr )
x {\rm d}x
\nonumber\\
&&\times
\biggl [ \sum_{n=0}^{\infty}\frac{\Gamma (n+\frac{3}{2})(n+1)}
{\Gamma(\frac{3}{2}n+2)} (-\chi x^{\frac{3}{2}})^{n}
\biggr ] .
\end{eqnarray}
Setting ${\bf q}={\bf 0}$ in Eq. (\ref{72}) and employing
normalization condition (\ref{20}), we infer that
\begin{equation}\label{73}
\bar{G} = \frac{\sqrt{\pi}}{4}
\biggl [ \sum_{n=0}^{\infty}\frac{\Gamma (n+\frac{3}{2})(n+1)}
{\Gamma(\frac{3}{2}n+3)} (-\chi)^{n} \biggr ]^{-1}.
\end{equation}
Taking into account that for any $x\geq 0$,
\begin{eqnarray}\label{74}
&& \sum_{n=0}^{\infty}
\frac{\Gamma(n+\frac{3}{2})(n+1)}{\Gamma(\frac{3}{2}n+2)}
(-x)^{n}
\nonumber\\
&& =\frac{\sqrt{\pi}}{2}
\int_{0}^{\infty}E_{\frac{3}{2},2}^{\frac{3}{2}}(-x y)
\exp(-y) y {\rm d}y ,
\end{eqnarray}
we derive an explicit presentation of the scattering function
in terms of integrals of the Mittag-Leffler function
\begin{eqnarray}\label{75}
S({\bf q}) &=& \int_{0}^{1} \exp \Bigl (-\frac{b^{2}q^{2}}{2}(1-x)\Bigr )
x {\rm d}x
\nonumber\\
&&\times
\frac{\int_{0}^{\infty}E_{\frac{3}{2},2}^{\frac{3}{2}}
(-\chi x^{\frac{3}{2}} y) \exp(-y) y {\rm d}y}
{\int_{0}^{\infty}E_{\frac{3}{2},2}^{\frac{3}{2}}
(-\chi y) \exp(-y) y {\rm d}y } .
\end{eqnarray}
To determine the radius of gyration $R_{\rm g}$,
we insert Eq. (\ref{72}) into Eq. (\ref{21}),
calculate the integrals and find that
\begin{equation}\label{76}
R_{\rm g}^{2}=\frac{2 b^{2}\bar{G}}{\sqrt{\pi}}
\sum_{n=0}^{\infty}\frac{\Gamma (n+\frac{3}{2})(n+1)}
{\Gamma(\frac{3}{2}n+4)} (-\chi)^{n}.
\end{equation}
By analogy with Eq. (\ref{74}), one can write
\begin{eqnarray}\label{77}
&& \sum_{n=0}^{\infty} \frac{\Gamma(n+\frac{3}{2})(n+1)}
{\Gamma(\frac{3}{2}n+4)} (-x)^{n}
\nonumber\\
&&=\frac{\sqrt{\pi}}{2}
\int_{0}^{\infty}E_{\frac{3}{2},4}^{\frac{3}{2}}(-x y)
\exp(-y) y {\rm d}y.
\end{eqnarray}
Combination of expressions (\ref{73}), (\ref{74}), (\ref{76})
and (\ref{77}) implies that
\begin{eqnarray}\label{78}
R_{\rm g}^{2} &=& b^{2}
\int_{0}^{\infty}E_{\frac{3}{2},4}^{\frac{3}{2}}(-\chi y)
\exp(-y) y {\rm d}y
\nonumber\\
&&\times
\biggl [\int_{0}^{\infty}E_{\frac{3}{2},2}^{\frac{3}{2}}(-\chi y)
\exp(-y) y {\rm d}y \biggr ]^{-1}.
\end{eqnarray}
At weak self-avoiding interactions, when the dimensionless
strength of segment interactions $\chi$ is small compared with
unity,
we disregard terms beyond the first order of smallness in
Eqs. (\ref{73}) and (\ref{77}) to obtain
\begin{equation}\label{79}
R_{\rm g}^{2}=\frac{b^{2}}{6}\Bigl (1+\frac{32\chi}{105\sqrt{\pi}}\Bigr ).
\end{equation}

\setlength{\unitlength}{0.55 mm}
\begin{figure}[t]
\begin{center}
\begin{picture}(100,100)
\put(0,0){\framebox(100,100)}
\multiput(12.5,0)(12.5,0){7}{\line(0,1){2}}
\multiput(0,10)(0,10){9}{\line(1,0){2}}
\put(0,-10){0.0}
\put(95,-10){8.0}
\put(50,-10){$\chi$}
\put(-11,0){1.0}
\put(-11,96){2.0}
\put(-11,70){$\xi$}

\put(   0.25,    0.34){\circle*{1.2}}
\put(   0.50,    0.69){\circle*{1.2}}
\put(   0.75,    1.03){\circle*{1.2}}
\put(   1.00,    1.37){\circle*{1.2}}
\put(   1.25,    1.71){\circle*{1.2}}
\put(   1.50,    2.06){\circle*{1.2}}
\put(   1.75,    2.40){\circle*{1.2}}
\put(   2.00,    2.74){\circle*{1.2}}
\put(   2.25,    3.08){\circle*{1.2}}
\put(   2.50,    3.42){\circle*{1.2}}
\put(   2.75,    3.76){\circle*{1.2}}
\put(   3.00,    4.10){\circle*{1.2}}
\put(   3.25,    4.43){\circle*{1.2}}
\put(   3.50,    4.77){\circle*{1.2}}
\put(   3.75,    5.11){\circle*{1.2}}
\put(   4.00,    5.45){\circle*{1.2}}
\put(   4.25,    5.78){\circle*{1.2}}
\put(   4.50,    6.12){\circle*{1.2}}
\put(   4.75,    6.45){\circle*{1.2}}
\put(   5.00,    6.79){\circle*{1.2}}
\put(   5.25,    7.12){\circle*{1.2}}
\put(   5.50,    7.46){\circle*{1.2}}
\put(   5.75,    7.79){\circle*{1.2}}
\put(   6.00,    8.12){\circle*{1.2}}
\put(   6.25,    8.46){\circle*{1.2}}
\put(   6.50,    8.79){\circle*{1.2}}
\put(   6.75,    9.12){\circle*{1.2}}
\put(   7.00,    9.45){\circle*{1.2}}
\put(   7.25,    9.78){\circle*{1.2}}
\put(   7.50,   10.11){\circle*{1.2}}
\put(   7.75,   10.44){\circle*{1.2}}
\put(   8.00,   10.77){\circle*{1.2}}
\put(   8.25,   11.10){\circle*{1.2}}
\put(   8.50,   11.43){\circle*{1.2}}
\put(   8.75,   11.75){\circle*{1.2}}
\put(   9.00,   12.08){\circle*{1.2}}
\put(   9.25,   12.41){\circle*{1.2}}
\put(   9.50,   12.73){\circle*{1.2}}
\put(   9.75,   13.06){\circle*{1.2}}
\put(  10.00,   13.38){\circle*{1.2}}
\put(  10.25,   13.70){\circle*{1.2}}
\put(  10.50,   14.03){\circle*{1.2}}
\put(  10.75,   14.35){\circle*{1.2}}
\put(  11.00,   14.67){\circle*{1.2}}
\put(  11.25,   14.99){\circle*{1.2}}
\put(  11.50,   15.31){\circle*{1.2}}
\put(  11.75,   15.63){\circle*{1.2}}
\put(  12.00,   15.95){\circle*{1.2}}
\put(  12.25,   16.27){\circle*{1.2}}
\put(  12.50,   16.59){\circle*{1.2}}
\put(  12.75,   16.91){\circle*{1.2}}
\put(  13.00,   17.23){\circle*{1.2}}
\put(  13.25,   17.54){\circle*{1.2}}
\put(  13.50,   17.86){\circle*{1.2}}
\put(  13.75,   18.18){\circle*{1.2}}
\put(  14.00,   18.49){\circle*{1.2}}
\put(  14.25,   18.80){\circle*{1.2}}
\put(  14.50,   19.12){\circle*{1.2}}
\put(  14.75,   19.43){\circle*{1.2}}
\put(  15.00,   19.74){\circle*{1.2}}
\put(  15.25,   20.06){\circle*{1.2}}
\put(  15.50,   20.37){\circle*{1.2}}
\put(  15.75,   20.68){\circle*{1.2}}
\put(  16.00,   20.99){\circle*{1.2}}
\put(  16.25,   21.30){\circle*{1.2}}
\put(  16.50,   21.61){\circle*{1.2}}
\put(  16.75,   21.91){\circle*{1.2}}
\put(  17.00,   22.22){\circle*{1.2}}
\put(  17.25,   22.53){\circle*{1.2}}
\put(  17.50,   22.84){\circle*{1.2}}
\put(  17.75,   23.14){\circle*{1.2}}
\put(  18.00,   23.45){\circle*{1.2}}
\put(  18.25,   23.75){\circle*{1.2}}
\put(  18.50,   24.05){\circle*{1.2}}
\put(  18.75,   24.36){\circle*{1.2}}
\put(  19.00,   24.66){\circle*{1.2}}
\put(  19.25,   24.96){\circle*{1.2}}
\put(  19.50,   25.26){\circle*{1.2}}
\put(  19.75,   25.56){\circle*{1.2}}
\put(  20.00,   25.86){\circle*{1.2}}
\put(  20.50,   26.46){\circle*{1.2}}
\put(  21.00,   27.06){\circle*{1.2}}
\put(  21.50,   27.65){\circle*{1.2}}
\put(  22.00,   28.24){\circle*{1.2}}
\put(  22.50,   28.83){\circle*{1.2}}
\put(  23.00,   29.42){\circle*{1.2}}
\put(  23.50,   30.00){\circle*{1.2}}
\put(  24.00,   30.58){\circle*{1.2}}
\put(  24.50,   31.16){\circle*{1.2}}
\put(  25.00,   31.73){\circle*{1.2}}
\put(  25.50,   32.31){\circle*{1.2}}
\put(  26.00,   32.88){\circle*{1.2}}
\put(  26.50,   33.45){\circle*{1.2}}
\put(  27.00,   34.01){\circle*{1.2}}
\put(  27.50,   34.57){\circle*{1.2}}
\put(  28.00,   35.13){\circle*{1.2}}
\put(  28.50,   35.69){\circle*{1.2}}
\put(  29.00,   36.25){\circle*{1.2}}
\put(  29.50,   36.80){\circle*{1.2}}
\put(  30.00,   37.35){\circle*{1.2}}
\put(  30.50,   37.90){\circle*{1.2}}
\put(  31.00,   38.44){\circle*{1.2}}
\put(  31.50,   38.98){\circle*{1.2}}
\put(  32.00,   39.52){\circle*{1.2}}
\put(  32.50,   40.06){\circle*{1.2}}
\put(  33.00,   40.60){\circle*{1.2}}
\put(  33.50,   41.13){\circle*{1.2}}
\put(  34.00,   41.66){\circle*{1.2}}
\put(  34.50,   42.19){\circle*{1.2}}
\put(  35.00,   42.71){\circle*{1.2}}
\put(  35.50,   43.24){\circle*{1.2}}
\put(  36.00,   43.76){\circle*{1.2}}
\put(  36.50,   44.27){\circle*{1.2}}
\put(  37.00,   44.79){\circle*{1.2}}
\put(  37.50,   45.30){\circle*{1.2}}
\put(  38.00,   45.81){\circle*{1.2}}
\put(  38.50,   46.32){\circle*{1.2}}
\put(  39.00,   46.83){\circle*{1.2}}
\put(  39.50,   47.33){\circle*{1.2}}
\put(  40.00,   47.83){\circle*{1.2}}
\put(  40.50,   48.33){\circle*{1.2}}
\put(  41.00,   48.83){\circle*{1.2}}
\put(  41.50,   49.32){\circle*{1.2}}
\put(  42.00,   49.81){\circle*{1.2}}
\put(  42.50,   50.30){\circle*{1.2}}
\put(  43.00,   50.79){\circle*{1.2}}
\put(  43.50,   51.27){\circle*{1.2}}
\put(  44.00,   51.76){\circle*{1.2}}
\put(  44.50,   52.24){\circle*{1.2}}
\put(  45.00,   52.71){\circle*{1.2}}
\put(  45.50,   53.19){\circle*{1.2}}
\put(  46.00,   53.66){\circle*{1.2}}
\put(  46.50,   54.13){\circle*{1.2}}
\put(  47.00,   54.60){\circle*{1.2}}
\put(  47.50,   55.07){\circle*{1.2}}
\put(  48.00,   55.53){\circle*{1.2}}
\put(  48.50,   56.00){\circle*{1.2}}
\put(  49.00,   56.46){\circle*{1.2}}
\put(  49.50,   56.91){\circle*{1.2}}
\put(  50.00,   57.37){\circle*{1.2}}
\put(  50.50,   57.82){\circle*{1.2}}
\put(  51.00,   58.27){\circle*{1.2}}
\put(  51.50,   58.72){\circle*{1.2}}
\put(  52.00,   59.17){\circle*{1.2}}
\put(  52.50,   59.62){\circle*{1.2}}
\put(  53.00,   60.06){\circle*{1.2}}
\put(  53.50,   60.50){\circle*{1.2}}
\put(  54.00,   60.94){\circle*{1.2}}
\put(  54.50,   61.37){\circle*{1.2}}
\put(  55.00,   61.81){\circle*{1.2}}
\put(  55.50,   62.24){\circle*{1.2}}
\put(  56.00,   62.67){\circle*{1.2}}
\put(  56.50,   63.10){\circle*{1.2}}
\put(  57.00,   63.53){\circle*{1.2}}
\put(  57.50,   63.95){\circle*{1.2}}
\put(  58.00,   64.37){\circle*{1.2}}
\put(  58.50,   64.79){\circle*{1.2}}
\put(  59.00,   65.21){\circle*{1.2}}
\put(  59.50,   65.63){\circle*{1.2}}
\put(  60.00,   66.04){\circle*{1.2}}
\put(  60.50,   66.46){\circle*{1.2}}
\put(  61.00,   66.87){\circle*{1.2}}
\put(  61.50,   67.28){\circle*{1.2}}
\put(  62.00,   67.68){\circle*{1.2}}
\put(  62.50,   68.09){\circle*{1.2}}
\put(  63.00,   68.49){\circle*{1.2}}
\put(  63.50,   68.89){\circle*{1.2}}
\put(  64.00,   69.29){\circle*{1.2}}
\put(  64.50,   69.69){\circle*{1.2}}
\put(  65.00,   70.08){\circle*{1.2}}
\put(  65.50,   70.48){\circle*{1.2}}
\put(  66.00,   70.87){\circle*{1.2}}
\put(  66.50,   71.26){\circle*{1.2}}
\put(  67.00,   71.65){\circle*{1.2}}
\put(  67.50,   72.03){\circle*{1.2}}
\put(  68.00,   72.42){\circle*{1.2}}
\put(  68.50,   72.80){\circle*{1.2}}
\put(  69.00,   73.18){\circle*{1.2}}
\put(  69.50,   73.56){\circle*{1.2}}
\put(  70.00,   73.94){\circle*{1.2}}
\put(  70.50,   74.32){\circle*{1.2}}
\put(  71.00,   74.69){\circle*{1.2}}
\put(  71.50,   75.06){\circle*{1.2}}
\put(  72.00,   75.44){\circle*{1.2}}
\put(  72.50,   75.80){\circle*{1.2}}
\put(  73.00,   76.17){\circle*{1.2}}
\put(  73.50,   76.54){\circle*{1.2}}
\put(  74.00,   76.90){\circle*{1.2}}
\put(  74.50,   77.27){\circle*{1.2}}
\put(  75.00,   77.63){\circle*{1.2}}
\put(  75.50,   77.99){\circle*{1.2}}
\put(  76.00,   78.34){\circle*{1.2}}
\put(  76.50,   78.70){\circle*{1.2}}
\put(  77.00,   79.05){\circle*{1.2}}
\put(  77.50,   79.41){\circle*{1.2}}
\put(  78.00,   79.76){\circle*{1.2}}
\put(  78.50,   80.11){\circle*{1.2}}
\put(  79.00,   80.46){\circle*{1.2}}
\put(  79.50,   80.80){\circle*{1.2}}
\put(  80.00,   81.15){\circle*{1.2}}
\put(  80.50,   81.49){\circle*{1.2}}
\put(  81.00,   81.84){\circle*{1.2}}
\put(  81.50,   82.18){\circle*{1.2}}
\put(  82.00,   82.52){\circle*{1.2}}
\put(  82.50,   82.85){\circle*{1.2}}
\put(  83.00,   83.19){\circle*{1.2}}
\put(  83.50,   83.53){\circle*{1.2}}
\put(  84.00,   83.86){\circle*{1.2}}
\put(  84.50,   84.19){\circle*{1.2}}
\put(  85.00,   84.52){\circle*{1.2}}
\put(  85.50,   84.85){\circle*{1.2}}
\put(  86.00,   85.18){\circle*{1.2}}
\put(  86.50,   85.51){\circle*{1.2}}
\put(  87.00,   85.83){\circle*{1.2}}
\put(  87.50,   86.15){\circle*{1.2}}
\put(  88.00,   86.48){\circle*{1.2}}
\put(  88.50,   86.80){\circle*{1.2}}
\put(  89.00,   87.12){\circle*{1.2}}
\put(  89.50,   87.43){\circle*{1.2}}
\put(  90.00,   87.75){\circle*{1.2}}
\put(  90.50,   88.07){\circle*{1.2}}
\put(  91.00,   88.38){\circle*{1.2}}
\put(  91.50,   88.69){\circle*{1.2}}
\put(  92.00,   89.01){\circle*{1.2}}
\put(  92.50,   89.32){\circle*{1.2}}
\put(  93.00,   89.63){\circle*{1.2}}
\put(  93.50,   89.93){\circle*{1.2}}
\put(  94.00,   90.24){\circle*{1.2}}
\put(  94.50,   90.54){\circle*{1.2}}
\put(  95.00,   90.85){\circle*{1.2}}
\put(  95.50,   91.15){\circle*{1.2}}
\put(  96.00,   91.45){\circle*{1.2}}
\put(  96.50,   91.75){\circle*{1.2}}
\put(  97.00,   92.05){\circle*{1.2}}
\put(  97.50,   92.35){\circle*{1.2}}
\put(  98.00,   92.65){\circle*{1.2}}
\put(  98.50,   92.95){\circle*{1.2}}
\put(  99.00,   93.24){\circle*{1.2}}
\put(  99.50,   93.53){\circle*{1.2}}
\put( 100.00,   93.83){\circle*{1.2}}
\put(   1.60,    2.75){\circle{1.8}}
\put(   3.20,    5.50){\circle{1.8}}
\put(   4.80,    8.25){\circle{1.8}}
\put(   6.40,   11.00){\circle{1.8}}
\put(   8.00,   13.76){\circle{1.8}}
\put(   9.60,   16.51){\circle{1.8}}
\put(  11.20,   19.26){\circle{1.8}}
\put(  12.80,   22.01){\circle{1.8}}
\put(  14.40,   24.76){\circle{1.8}}
\put(  16.00,   27.51){\circle{1.8}}
\put(  17.60,   30.26){\circle{1.8}}
\put(  19.20,   33.01){\circle{1.8}}
\put(  20.80,   35.76){\circle{1.8}}
\put(  22.40,   38.52){\circle{1.8}}
\put(  24.00,   41.27){\circle{1.8}}
\put(  25.60,   44.02){\circle{1.8}}
\put(  27.20,   46.77){\circle{1.8}}
\put(  28.80,   49.52){\circle{1.8}}
\put(  30.40,   52.27){\circle{1.8}}
\put(  32.00,   55.02){\circle{1.8}}
\put(  33.60,   57.77){\circle{1.8}}
\put(  35.20,   60.52){\circle{1.8}}
\put(  36.80,   63.28){\circle{1.8}}
\put(  38.40,   66.03){\circle{1.8}}
\put(  40.00,   68.78){\circle{1.8}}
\put(  41.60,   71.53){\circle{1.8}}
\put(  43.20,   74.28){\circle{1.8}}
\put(  44.80,   77.03){\circle{1.8}}
\put(  46.40,   79.78){\circle{1.8}}
\put(  48.00,   82.53){\circle{1.8}}
\put(  49.60,   85.28){\circle{1.8}}
\put(  51.20,   88.04){\circle{1.8}}
\put(  52.80,   90.79){\circle{1.8}}
\put(  54.40,   93.54){\circle{1.8}}
\put(  56.00,   96.29){\circle{1.8}}
\put(  57.60,   99.04){\circle{1.8}}

\end{picture}
\end{center}
\vspace*{5 mm}

\caption{
The dimensionless ratio $\xi$
versus the dimensionless strength of segment interactions
$\chi$.
Solid line: exact solution.
Circles: linear approximation.
\label{f-1}}
\end{figure}
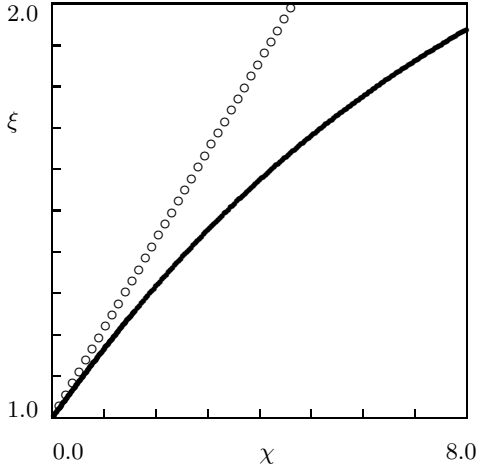

Equation (\ref{79}) shows that at any $\chi >0$,
the radius of gyration of a weakly self-repellent chain
exceeds that of a standard Gaussian chain,
$R_{\rm g\; 0}^{2}=\frac{1}{6}b^{2}$.
To evaluate the effect of excluded-volume interactions
on the radius of gyration, we perform summation of the
infinite series (\ref{73}) and (\ref{76}) numerically.
The dimensionless ratio
$\xi=R_{\rm g}^{2}/R_{\rm g\;0}^{2}$
is plotted versus the dimensionless strength of segment
interactions $\chi$ in Figure 1, which shows that
the radius of gyration increases monotonically with strength
of self-repellent interactions.
Linear approximation (\ref{79}) ensures a correct fit of
the dependence $R_{\rm g}(\chi)$ at relatively
small strengths ($\chi<1)$, but overly estimates the radius
of gyration at higher values of $\chi$.

\section{A self-avoiding chain in a three-dimensional space}

For a self-repellent chain in a three-dimensional space
($d=3$), the use of a regularizing function $\rho(t)$ is necessary,
because the integral with $\rho=1$ in Eq. (\ref{51})
diverges.
It follows from Eqs. (\ref{5}) and (\ref{51}) that
\[
\Lambda(\lambda)=
2\sqrt{\pi}\Bigl (\sqrt{\lambda+r}-\sqrt{\lambda}\Bigr ),
\]
where $r=L_{0}^{-1}$.
This relation together with Eq. (\ref{62}) results in
\begin{eqnarray}\label{80}
W(\lambda,z)&=& \frac{2\sqrt{\pi}}{\lambda^{2}}W_{1}(\lambda,z),
\quad
\phi(z)=4\sqrt{\pi} v_{0}z,
\nonumber\\
W_{1}(\lambda,z)&=&
\frac{\sqrt{\lambda+r}-\sqrt{\lambda}}
{\lambda+\phi(z) (\sqrt{\lambda+r}-\sqrt{\lambda})}.
\end{eqnarray}
Bearing in mind that for any $X$,
\[
\frac{X}{1+\phi X}=\sum_{n=1}^{\infty} (-\phi)^{n-1}X^{n},
\]
we infer that
\begin{equation}\label{81}
W_{1}(\lambda,z)=\sum_{n=1}^{\infty} (-\phi)^{n-1}
\Bigl ( \frac{\sqrt{\lambda+r}-\sqrt{\lambda}}{\lambda}\Bigr )^{n}.
\end{equation}
To find the inverse Laplace transform of the function $W(\lambda,z)$,
we begin with the formula
\begin{equation}\label{82}
{\cal L}^{-1}\Bigl ((\sqrt{s^{2}+a}-s)^{n}\Bigr )_{s\to x}
=n a^{\frac{n}{2}} \frac{J_{n}(\sqrt{a} x)}{x},
\end{equation}
where
\begin{equation}\label{83}
J_{n}(x)=\sum_{m=0}^{\infty}
\frac{(-1)^{m}}{m!\Gamma(m+n+1)}
\Bigl (\frac{x}{2}\Bigr )^{2m+n}
\end{equation}
is the Bessel function of the first kind of order $n$,
and $a>0$ is an arbitrary constant.
We now apply the well-known assertion which says
that if $\Phi(s)$ is the Laplace transform of a function $\phi(x)$,
then
\[
{\cal L}^{-1}\biggl (\frac{\Phi(\sqrt{s})}{\sqrt{s}}\biggr )_{s\to x}
=\frac{1}{\sqrt{\pi x}} \int_{0}^{\infty}
\phi(z)\exp \Bigl (-\frac{z^{2}}{4x}\Bigr ){\rm d}z.
\]
Applying this proposition to Eq. (\ref{82}), we obtain
\begin{eqnarray}\label{84}
&& {\cal L}^{-1}\Bigl (\frac{(\sqrt{s+a}-\sqrt{s})^{n}}{\sqrt{s}}
\Bigr )_{s\to x}
\nonumber\\
&& =\frac{n a^{\frac{n}{2}}}{\sqrt{\pi x}}
\int_{0}^{\infty} J_{n}(\sqrt{a} z)
\exp \Bigl (-\frac{z^{2}}{4x}\Bigr )z^{-1}{\rm d}z.
\end{eqnarray}
Inserting expression (\ref{83}) into Eq. (\ref{84}),
performing integration term by term,
and replacing the resulting series with an appropriate
Mittag--Leffler function, we infer that
\begin{eqnarray}\label{85}
&& {\cal L}^{-1}\Bigl (\frac{(\sqrt{s+a}-\sqrt{s})^{n}}{\sqrt{s}}
\Bigr )_{s\to x}
\nonumber\\
&&=\frac{n a^{n} x^{\frac{n-1}{2}}}{2 \sqrt{\pi}}
\Gamma \Bigl (\frac{n}{2}\Bigr )
E_{1,n+1}^{\frac{n}{2}}(-a x).
\end{eqnarray}
Bearing in mind that
\begin{eqnarray*}
&& {\cal L}^{-1}\Bigl ( \Bigl (\frac{\sqrt{s+a}-\sqrt{s}}{s}\Bigr )^{n}
\Bigr )_{s\to x}
\nonumber\\
&&
={\cal L}^{-1} \Bigl ( \frac{1}{s^{n-\frac{1}{2}}}
\frac{(\sqrt{s+a}-\sqrt{s})^{n}}{\sqrt{s}}\Bigr )_{s\to x},
\nonumber\\
&& {\cal L}^{-1} \Bigl ( \frac{1}{s^{n-\frac{1}{2}}}
\Bigr )_{s\to x}=\frac{x^{n-\frac{3}{2}}}{\Gamma(n-\frac{1}{2})},
\end{eqnarray*}
and applying the convolution theorem for the inverse
Laplace transform, we find that
\begin{eqnarray*}
&& {\cal L}^{-1}\Bigl ( \Bigl (\frac{\sqrt{s+a}-\sqrt{s}}{s}\Bigr )^{n}
\Bigr )_{s\to x}
=\frac{1}{\Gamma(n-\frac{1}{2})}
\nonumber\\
&&\times
\int_{0}^{x} (x-t)^{n-\frac{3}{2}}
{\cal L}^{-1} \Bigl (\frac{(\sqrt{s+a}-\sqrt{s})^{n}}{\sqrt{s}}
\Bigr )_{s\to t} {\rm d}t.
\end{eqnarray*}
Substitution of expression (\ref{85}) into this relation yields
\begin{eqnarray}\label{86}
&& {\cal L}^{-1}\Bigl ( \Bigl (\frac{\sqrt{s+a}-\sqrt{s}}{s}\Bigr )^{n}
\Bigr )_{s\to x}
=\frac{n a^{n}}{2 \sqrt{\pi}}
\frac{\Gamma(\frac{n}{2})}{\Gamma(n-\frac{1}{2})}
\nonumber\\
&& \times
\int_{0}^{x} (x-t)^{n-\frac{3}{2}}t^{\frac{n-1}{2}}
E_{1,n+1}^{\frac{n}{2}}(-a t){\rm d}t.
\end{eqnarray}
Combination of Eqs. (\ref{81}) and (\ref{86}) results in
\begin{eqnarray*}
&& {\cal L}^{-1} \Bigl ( W_{1}(\lambda,z)\Bigr )_{\lambda\to l}
= \frac{r\sqrt{l}}{2\sqrt{\pi}}
\sum_{n=1}^{\infty} \frac{n\Gamma(\frac{n}{2})}{\Gamma(n-\frac{1}{2})}
(-r l^{\frac{3}{2}} \phi)^{n-1}
\nonumber\\
&&\times
\int_{0}^{1} x^{\frac{n-1}{2}} (1-x)^{n-\frac{3}{2}}
E_{1,n+1}^{\frac{n}{2}}(-r l x){\rm d} x,
\end{eqnarray*}
where we set $x=t/l$.
Using the integral presentation of the Mittag--Leffler
function
\begin{eqnarray*}
&& E_{1,n+1}^{\frac{n}{2}}(-r l x)
=\frac{1}{\Gamma(\frac{n}{2})\Gamma(\frac{n}{2}+1)}
\nonumber\\
&&\times
\int_{0}^{1} y^{\frac{n}{2}-1}(1-y)^{\frac{n}{2}}
\exp(-r l x y) {\rm d}y ,
\end{eqnarray*}
we arrive at the formula
\begin{eqnarray}\label{87}
&&{\cal L}^{-1} \Bigl ( W_{1}(\lambda,z)\Bigr )_{\lambda\to l}
= \frac{r\sqrt{l} }{2\sqrt{\pi}}
\sum_{m=0}^{\infty} \frac{(m+1)(-r l^{\frac{3}{2}} \phi)^{m}}
{\Gamma(m+\frac{1}{2})\Gamma(\frac{m}{2}+\frac{3}{2})}
\nonumber\\
&&\times
\int_{0}^{1} x^{\frac{m}{2}} (1-x)^{m-\frac{1}{2}}
{\rm d} x
\nonumber\\
&&\times
\int_{0}^{1} y^{\frac{m-1}{2}}(1-y)^{\frac{m+1}{2}}
\exp(-r l x y) {\rm d}y .
\end{eqnarray}
Application of the convolution theorem for the inverse
Laplace transform to Eq. (\ref{80}) results in
\begin{eqnarray}\label{88}
&&{\cal L}^{-1}\Bigl (W(\lambda,z)\Bigr )_{\lambda\to l}
\nonumber\\
&&=
2\sqrt{\pi} \int_{0}^{l} (l-t) {\cal L}^{-1}\Bigl (W_{1}(\lambda,z)
\Bigr )_{\lambda\to t} {\rm d}t,
\end{eqnarray}
where we employed the formula
${\cal L}^{-1}(\lambda^{-2})_{\lambda\to x}=x$.
Setting $u=t/l$ in Eq. (\ref{88}) and substituting
expression (\ref{87}), we find that
\begin{eqnarray}\label{89}
&&{\cal L}^{-1} \Bigl ( W(\lambda,z)\Bigr )_{\lambda\to l}
= r l^{\frac{5}{2}} \sum_{m=0}^{\infty}
\frac{(m+1)(-r l^{\frac{3}{2}} \phi)^{m}}
{\Gamma(m+\frac{1}{2})\Gamma(\frac{m}{2}+\frac{3}{2})}
\nonumber\\
&&\times
\int_{0}^{1} u^{\frac{3m+1}{2}} (1-u) {\rm d}u
\int_{0}^{1} x^{\frac{m}{2}} (1-x)^{m-\frac{1}{2}}
{\rm d} x
\nonumber\\
&&\times
\int_{0}^{1} y^{\frac{m-1}{2}}(1-y)^{\frac{m+1}{2}}
\exp(-r l u x y) {\rm d}y .
\end{eqnarray}
Combination of Eqs. (\ref{61}) and (\ref{89}) yields
\begin{widetext}
\begin{eqnarray*}
P(L,{\bf q}) &=&  \frac{4v_{0}r }{\sqrt{\pi}}
\sum_{m=0}^{\infty}
\frac{(m+1)(-r)^{m}}{\Gamma(m+\frac{1}{2})
\Gamma(\frac{m}{2}+\frac{3}{2})}
\int_{0}^{\infty} \exp(-z)(1-2z) z^{\frac{3}{2}} \phi^{m}(z)
{\rm d}z
\int_{0}^{L} \exp \Bigl (-\frac{b_{0}(L-l)q^{2}}{6}\Bigr )
l^{\frac{3m+5}{2}} {\rm d}l
\nonumber\\
&&\times
\int_{0}^{1} u^{\frac{3m+1}{2}} (1-u) {\rm d}u
\int_{0}^{1} x^{\frac{m}{2}} (1-x)^{m-\frac{1}{2}}
{\rm d} x
\int_{0}^{1} y^{\frac{m-1}{2}}(1-y)^{\frac{m+1}{2}}
\exp(-r l u x y) {\rm d}y .
\end{eqnarray*}
Introducing the notation $t=l/L$ and inserting expression (\ref{13}),
(\ref{54}) and (\ref{80}) into this relation, we find that
\begin{eqnarray}\label{90}
P(L,{\bf q}) &=&  \frac{2\bar{\chi} L^{2}}{\pi}
\sum_{m=0}^{\infty}
\frac{(-\bar{\chi})^{m}}{\Gamma(m+\frac{1}{2})
\Gamma(\frac{m}{2}+\frac{1}{2})}
\int_{0}^{\infty} \exp(-z)(1-2z) z^{m+\frac{3}{2}}{\rm d}z
\int_{0}^{1} \exp \Bigl (-\frac{b^{2}q^{2}}{6}(1-t)\Bigr )
t^{\frac{3m+5}{2}} {\rm d}t
\nonumber\\
&&\times
\int_{0}^{1} u^{\frac{3m+1}{2}} (1-u) {\rm d}u
\int_{0}^{1} x^{\frac{m}{2}} (1-x)^{m-\frac{1}{2}}
{\rm d} x
\int_{0}^{1} y^{\frac{m-1}{2}}(1-y)^{\frac{m+1}{2}}
\exp(-\kappa t u x y) {\rm d}y ,
\end{eqnarray}
\end{widetext}
where
\begin{equation}\label{91}
\bar{\chi}=\frac{3 \kappa v}{\pi} \sqrt{\frac{6b_{0}}{L}}.
\end{equation}
The integral over $z$ is calculated explicitly by means of
the gamma-function,
\begin{equation}\label{92}
\int_{0}^{\infty} \exp(-z)(1-2z) z^{m+\frac{3}{2}}{\rm d}z
=-2(m+2)\Gamma\Bigl (m+\frac{5}{2}\Bigr ).
\end{equation}
Inserting expressions (\ref{46}) and (\ref{90})
into Eq. (\ref{47}) and using Eq. (\ref{92})
and elementary properties of the gamma-function,
we arrive at the formula for the scattering function
of a self-avoiding chain in a three-dimensional space
\begin{eqnarray}\label{93}
&& S({\bf q}) =  2\bar{G}
\int_{0}^{1} \exp \Bigl (-\frac{b^{2}q^{2}}{6}(1-t)\Bigr )
t {\rm d}t
\biggl [ 1
\nonumber\\
&& +\frac{2}{\pi} \sum_{m=0}^{\infty}
\frac{(m+\frac{1}{2})(m+\frac{3}{2})(m+2)
(-\bar{\chi} t^{\frac{3}{2}})^{m+1}}
{\Gamma(\frac{m}{2}+\frac{1}{2})}
\nonumber\\
&&\times
\int_{0}^{1} u^{\frac{3m+1}{2}} (1-u) {\rm d}u
\int_{0}^{1} x^{\frac{m}{2}} (1-x)^{m-\frac{1}{2}}
{\rm d} x
\nonumber\\
&&\times
\int_{0}^{1} y^{\frac{m-1}{2}}(1-y)^{\frac{m+1}{2}}
\exp(-\kappa t u x y) {\rm d}y \biggr ] .
\end{eqnarray}
The function (\ref{93}) is determined by three parameters:
(i) the mean-square-root end-to-end distance of
an appropriate Gaussian chain $b$,
(ii) the dimensionless strength of segment interactions $\bar{\chi}$,
and (iii) the dimensionless length of a chain $\kappa$.
The pre-factor $\bar{G}$ is found from normalization
condition (\ref{20}).

\subsection{Weak excluded-volume interactions}

At weak self-avoiding interactions between segments
($\bar{\chi}\ll 1$),
we disregard all terms beyond the first order of smallness
in Eq. (\ref{93}) and obtain
\begin{eqnarray}\label{94}
S({\bf q}) &=&  2 \bar{G}
\int_{0}^{1} \exp \Bigl (-\frac{b^{2}q^{2}}{6}(1-t)\Bigr )
t {\rm d}t
\nonumber\\
&&\times
\biggl [ 1 -\frac{3\bar{\chi}t^{\frac{3}{2}}}{\pi\sqrt{\pi}}
\int_{0}^{1} u^{\frac{1}{2}} (1-u) {\rm d}u
\int_{0}^{1} (1-x)^{-\frac{1}{2}}{\rm d} x
\nonumber\\
&&\times
\int_{0}^{1} y^{-\frac{1}{2}}(1-y)^{\frac{1}{2}}
\exp(-\kappa t u x y) {\rm d}y \biggr ].
\end{eqnarray}
It follows from Eq. (\ref{66}) that
\[
\int_{0}^{1}y^{-\frac{1}{2}}(1-y)^{\frac{1}{2}}
\exp(-\kappa t u x y) {\rm d}y
=\frac{\pi}{2}E_{1,2}^{\frac{1}{2}}(-\kappa t u x).
\]
Taking into account that for an arbitrary positive $z$,
\[
\int_{0}^{1} (1-x)^{-\frac{1}{2}} E_{1,2}^{\frac{1}{2}}(-x z){\rm d} x
= \sum_{n=0}^{\infty}
\frac{\Gamma(n+\frac{1}{2})(-z)^{n}}{\Gamma(n+\frac{3}{2})\Gamma(n+2)},
\]
we find from Eq. (\ref{94}) that
\begin{eqnarray*}
&& S({\bf q})= 2 \bar{G}
\int_{0}^{1} \exp \Bigl (-\frac{b^{2}q^{2}}{6}(1-t)\Bigr )
t {\rm d}t
\biggl [ 1
\nonumber\\
&&-\frac{3\bar{\chi}t^{\frac{3}{2}}}{2\sqrt{\pi}}
\sum_{n=0}^{\infty} \frac{\Gamma(n+\frac{1}{2})(-\kappa t)^{n}}
{\Gamma(n+2)\Gamma(n+\frac{3}{2})}
\int_{0}^{1} u^{n+\frac{1}{2}} (1-u) {\rm d}u \biggr ].
\end{eqnarray*}
Calculating the internal integral, we arrive at the formula
\begin{eqnarray}\label{95}
&& S({\bf q}) = 2 \bar{G} \int_{0}^{1}
\exp \Bigl (-\frac{b^{2}q^{2}}{6}(1-t)\Bigr )
\nonumber\\
&&\times
\biggl [ 1 -\frac{3\bar{\chi}t^{\frac{3}{2}}}{2\sqrt{\pi}}
\sum_{n=0}^{\infty} \frac{\Gamma(n+\frac{1}{2})(-\kappa t)^{n}}
{\Gamma(n+2)\Gamma(n+\frac{7}{2})} \biggr ] t {\rm d}t .
\end{eqnarray}
To determine the coefficient $\bar{G}$, we set ${\bf q}= {\bf 0}$
in Eq. (\ref{95}), substitute the obtained expression into
Eq. (\ref{20}), perform integration, and obtain
\begin{equation}\label{96}
\bar{G} = \biggl [ 1 -\frac{3\bar{\chi}}{\sqrt{\pi}}
\sum_{n=0}^{\infty} \frac{\Gamma(n+\frac{1}{2})(-\kappa)^{n}}
{\Gamma(n+2)\Gamma(n+\frac{9}{2})} \biggr ]^{-1} .
\end{equation}
It follows from Eqs. (\ref{23}) and (\ref{95}) that
\begin{equation}\label{97}
R_{\rm g}^{2}
= \frac{b^{2}\bar{G}}{6} \biggl [ 1
-\frac{9\bar{\chi}}{\sqrt{\pi}}
\sum_{n=0}^{\infty} \frac{\Gamma(n+\frac{1}{2})(-\kappa )^{n}}
{\Gamma(n+2)\Gamma(n+\frac{11}{2})} \biggr ] .
\end{equation}
where the integrals are calculated term by term.
Combining Eqs. (\ref{95}) and (\ref{96}) and neglecting
terms beyond the first order of smallness with respect to
$\bar{\chi}$, we find that
\begin{equation}\label{98}
R_{\rm g}^{2} = \frac{b^{2}}{6} \biggl [ 1
+\frac{3\bar{\chi}}{\sqrt{\pi}}
\sum_{n=0}^{\infty}
\frac{\Gamma(n+\frac{1}{2})(n+\frac{3}{2})(-\kappa )^{n}}
{\Gamma(n+2)\Gamma(n+\frac{11}{2})} \biggr ] .
\end{equation}
Bearing in mind that
\[
\frac{\Gamma(n+\frac{1}{2})(n+\frac{3}{2})}
{\Gamma(n+2)\Gamma(n+\frac{11}{2})}
=\frac{\Gamma(n+\frac{3}{2})+\Gamma(n+\frac{1}{2})}
{(n+1)\Gamma(n+\frac{11}{2})  n!}
\]
and
\[
\int_{0}^{1} E_{\alpha,\beta}^{\gamma}(-\kappa x) {\rm d}x
=\frac{1}{\Gamma(\gamma)}\sum_{n=0}^{\infty}
\frac{\Gamma(n+\gamma)}{(n+1)\Gamma(\alpha n+\beta)}
\frac{(-\kappa)^{n}}{n!},
\]
we present Eq. (\ref{97}) in the form
\begin{equation}\label{99}
R_{\rm g}^{2} = \frac{b^{2}}{6} \biggl \{ 1
+3\bar{\chi} \int_{0}^{1} \Bigl [
E_{1,\frac{11}{2}}^{\frac{1}{2}}(-\kappa x)
+\frac{1}{2} E_{1,\frac{11}{2}}^{\frac{3}{2}}(-\kappa x)
\Bigr ] {\rm d}x \biggr \}.
\end{equation}
Equation (\ref{99}) provides an explicit expression
for the radius of gyration of a self-repellent chain
in the limit of small strengths of segment interactions
($\bar{\chi}\ll 1$) for an arbitrary dimensionless
parameter $\kappa$.

\subsection{Asymptotic expression for the scattering function}

Our aim now is to derive a formula for the scattering function
at strong self-repellent interactions with $v\to \infty$
under the assumption that $\kappa\to 0$ and the product
$v\kappa$ in Eq. (\ref{91}) remains finite,
\begin{equation}\label{100}
\lim_{v\to \infty,\; \kappa\to 0}
\frac{3 v \kappa}{\pi} \sqrt{\frac{6b_{0}}{L}} =2\chi >0.
\end{equation}
The coefficient ``2" is introduced in Eq. (\ref{100}) for
convenience of further transformations.

Replacing the exponential function in Eq. (\ref{93}) by unity
and calculating the integrals in terms of the gamma-function,
we obtain
\begin{eqnarray*}
&& S({\bf q})
=  \frac{4\bar{G}}{\pi}
\int_{0}^{1} \exp \Bigl (-\frac{b^{2}q^{2}}{6}(1-t)\Bigr )
t {\rm d}t
\nonumber\\
&&\times
\sum_{n=0}^{\infty}
\frac{(n+1)\Gamma(\frac{1}{2} n+\frac{1}{2})
\Gamma(\frac{1}{2}n+1)
\Gamma(n+\frac{3}{2})(-2 \chi t^{\frac{3}{2}})^{n}}
{\Gamma(n+1)\Gamma(\frac{3}{2}n+2)},
\end{eqnarray*}
where we included the first term in the square brackets into
the infinite sum.
Applying the Legendre duplication formula,
we find from this relation that
\begin{eqnarray}\label{101}
S({\bf q}) &=& \frac{4 \bar{G}}{\sqrt{\pi} }
\int_{0}^{1} \exp \Bigl (-\frac{b^{2}q^{2}}{6}(1-t)\Bigr )
t {\rm d}t
\nonumber\\
&&\times
\biggl [ \sum_{n=0}^{\infty} \frac{\Gamma(n+\frac{3}{2})(n+1)}
{\Gamma(\frac{3}{2}n+2)}(-\chi t^{\frac{3}{2}})^{n}
\biggr ]  .
\end{eqnarray}
Comparison of Eqs. (\ref{72}) and (\ref{101}) implies
that in the limit (\ref{100}), the scattering function
for a self-repellent chain in a three-dimensional space
coincides (after the natural replacement $\frac{1}{6} q^{2}
\to \frac{1}{2} q^{2}$) with that for a self-avoiding chain
in an one-dimensional space.
Based on this similarity and using Eq. (\ref{75}),
we arrive at the closed-form expression for the
scattering function
\begin{eqnarray}\label{102}
S({\bf q}) &=&
\int_{0}^{1} \exp \Bigl(-\frac{1}{6} b^{2}q^{2}(1-t)\Bigr )
t {\rm d}t
\nonumber\\
&&\times
\frac{\int_{0}^{\infty}E_{\frac{3}{2},2}^{\frac{3}{2}}
(-\chi t^{\frac{3}{2}} y) \exp(-y) y {\rm d}y}
{\int_{0}^{\infty}E_{\frac{3}{2},2}^{\frac{3}{2}}
(-\chi y) \exp(-y) y {\rm d}y }
\end{eqnarray}
that extends the Debye formula for a standard Gaussian chain
to self-avoiding chains.
It follows from Eqs. (\ref{23}) and (\ref{102}) that
the radius of gyration of a self-repellent chain in a
three-dimensional space is given by Eq. (\ref{78}),
and its dependence on strength of self-avoiding interactions
is presented in Figure 1.

\begin{figure}[tbh]
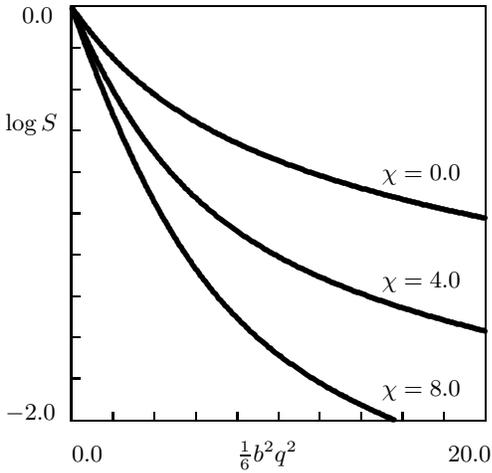

\begin{center}

\end{center}
\vspace*{5 mm}

\caption{
The Guinier plots of the scattering function $S$
for a standard Gaussian chain ($\chi=0.0$)
and for self-repellent chains with various strengths of
excluded-volume interactions.
\label{f-2}}
\end{figure}

\begin{figure}[tbh]
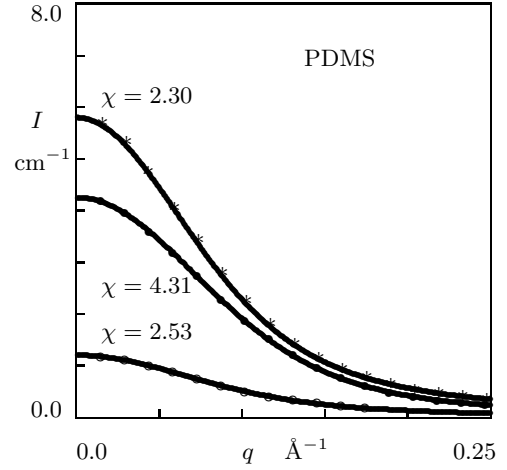

\begin{center}
%
\end{center}
\vspace*{5 mm}

\caption{
The scattering intensity $I$ versus
the amplitude of scattering vector $q$.
Symbols: experimental data on a PDMS blend
\cite{AGD04}
at volume fractions of H-PDMS 0.06 (unfilled circles),
0.26 (filled circles),
and 0.52 (asterisks).
Solid lines: results of numerical simulation.
\label{f-3}}
\end{figure}

\begin{figure}[tbh]
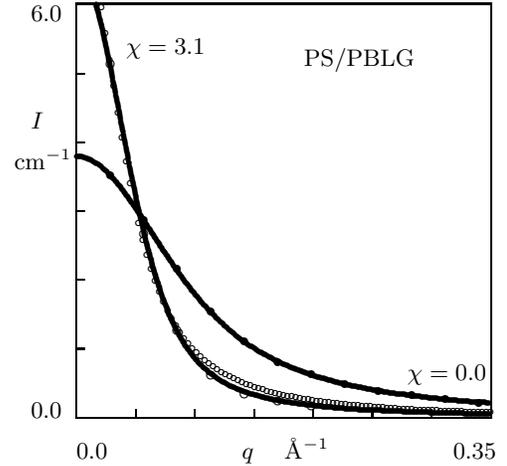

\begin{center}
%
\end{center}
\vspace*{5 mm}

\caption{
The scattering intensity $I$ versus the amplitude
of scattering vector $q$.
Large symbols: experimental data on solutions of PS/PBLG
\cite{CLB03}
in dioxane (unfilled circles)
and dioxane/TFA mixture (filled circles).
Solid lines: results of numerical simulation.
Small circles: approximation by the Debay function
with $b=137.2$ nm.
\label{f-4}}
\end{figure}

\begin{figure}[tbh]
\begin{center}
%
\end{center}
\vspace*{5 mm}

\caption{
The scattering intensity $I$ versus the
amplitude of scattering vector $q$.
Symbols: experimental data on a PDME/PEMS blend
at the temperatures $T=100$, 200 and 250~$^{\circ}$C,
from top to bottom, respectively \cite{YMW04}.
Solid lines: results of numerical simulation
with $\chi=0.84\;$.
\label{f-5}}
\end{figure}

\begin{figure}[tbh]
\begin{center}
%
\end{center}
\vspace*{5 mm}

\caption{
The scattering intensity $I$ versus the
amplitude of scattering vector $q$.
Symbols: experimental data on a PDME/PEMS blend
at the temperatures $T=150$, 200 and 300~$^{\circ}$C,
from top to bottom, respectively \cite{YMW04}.
Solid lines: results of numerical simulation
with $\chi=0.67\;$.
\label{f-6}}
\end{figure}

\begin{figure}[tbh]
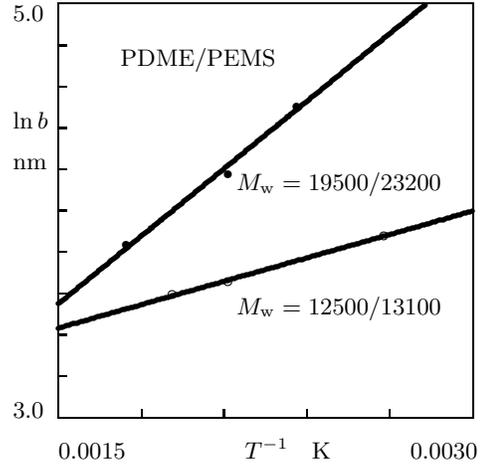

\begin{center}
%
\end{center}
\vspace*{5 mm}

\caption{
The mean-square-root end-to-end distance
for an equivalent Gaussian chain $b$ versus temperature $T$.
Symbols: treatment of observations on PDMS/PEMS blends.
Solid lines: approximation of the experimental data
by Eq. (104).
\label{f-7}}
\end{figure}

\section{Results of numerical simulation}

To perform numerical simulation, we return to
presentation (\ref{101}) of the scattering function
and write
\begin{eqnarray}\label{103}
S({\bf q}) &=& \Bigl (\sum_{n=0}^{\infty} c_{n}(-\chi)^{n}\Bigr )^{-1}
\int_{0}^{1} \exp \Bigl (-\frac{1}{6} b^{2}q^{2}(1-t)\Bigr )
\nonumber\\
&&\times
\sum_{n=0}^{\infty} a_{n}(-\chi t^{\frac{3}{2}})^{n}
t {\rm d}t ,
\end{eqnarray}
where
\[
a_{n}= \frac{(n+1)\Gamma(n+\frac{3}{2})}
{\Gamma(\frac{3}{2}n+2)},
\qquad
c_{n}=\frac{2a_{n}}{3n+4}.
\]
We sum numerically series (\ref{103}) for several values
of $\chi$ and present the dependence $S(q)$ in Figure 2
(the Guinier plot).
Given an amplitude of the scattering vector $q$, an increase
in strength of self-repellent interactions results in
a pronounced decrease in the scattering intensity.
With the growth of $\chi$,
the graph of the scattering function flattens at relatively
small values of $q$ and it becomes more close to the
graph of the Giunier function.
The latter function is believed to adequately describe
the scattering intensity of chains with strong excluded-volume
interactions \cite{CMC02}.

We proceed with matching observations
[small angle neutron scattering (SANS)]
on poly(dimethylsiloxane) (PDMS)
blend 5HC/11DC with various volume fractions of hydrogenated
PDMS (H-PDMS) reported in \cite{AGD04}.
A description of the experimental procedure is given in
\cite{AGD04}.
Assuming the scattering intensity $I(q)$ to be proportional
to the scattering function $S(q)$, $I=KS$, where $K$ is
an optical constant \cite{HB94}, we approximate
the experimental data depicted in Figure 3
by using three adjustable parameters: $b$, $\chi$ and $K$.
The best-fit quantities $b$ and $\chi$ are determined by
the steepest-descent method, and $K$ is found by
the least-squares algorithm.
Figure 3 demonstrates an excellent quality of
matching the observations with $\chi$ varying in the
interval between 2.3 and 4.3$\;$.

To compare the accuracy of fitting experimental data
by Eq. (\ref{103}) and by the Debye function (\ref{46}),
we approximate observations (SANS) on
poly(styrene-d$_{8}$)-b-poly(-benzyl L-glutamate)
(PS/PBLG) reported in \cite{CLB03}.
The scattering intensity $I$ is plotted versus the amplitude
of scattering vector $q$ in Figure 4
for solutions of PS/PBLG in dioxane and
mixture of dioxane and trifluoroacetic acid (TFA)
with 20 wt.-\% of TFA.
For a detailed description of the experimental procedure,
see \cite{CLB03}, where it is demonstrated that the presence
of TFA in the solution results in transition from the rod-coil
architecture of the diblock polymer to coil-coil topology
of macromolecules.
Figure 4 shows that the best-fit parameter $\chi$
vanishes for the coil-coil system, and Eq. (\ref{103})
is reduced to the Debye function.
For the rod-coil architecture of the diblock
polymer, $\chi$ equals 3.1 (approximately), which means that
the effect of excluded-volume interactions between rods
is substantial.
The Debye fit of the observations on solution of PS/PBLG in dioxane
is rather poor compared with the approximation by Eq. (\ref{103}),
and the best-fit value of $b$ found when self-avoiding
interactions are disregarded appears to be too large.

To study the effect of temperature $T$ and mass-average molecular
weight $M_{\rm w}$ on the material constants in Eq. (\ref{103}),
we approximate observations (SANS) on solutions
of symmetric blends of poly(dimethylsiloxane)
and poly(ethylmethylsiloxane) (PEMS) in toluene.
A detailed description of polymers and the experimental
procedure is given in \cite{YMW04}.
The experimental data and the results of numerical simulation
are presented in Figures 5 and 6 for two blends
at temperatures ranged from 100 to 300~$^{\circ}$C.
To reduce the number of adjustable parameters, we find $\chi$
for each blend by matching observations at the lowest
temperature and use this quantity without changes to
approximate the scattering intensity measured at other temperatures
(this means that $b$ is the only adjustable parameter in the fitting
procedure).
Figures 5 and 6 demonstrate good agreement between the observations
and the results of numerical analysis
(in contrast with the model used in the original work \cite{YMW04}
that contains the same number of adjustable constants, but shows
rather poor quality of matching the experimental data).
The parameter $\chi$ equals 0.84 for the blend with low molecular
weight and 0.67 for that with high $M_{\rm w}$.

The mean-square-root end-to-end distance for an equivalent
Gaussian chain $b$ determined by fitting these data
is depicted in Figure 7 together with its approximation by
the Arrhenius-type dependence
\begin{equation}\label{104}
\ln b=\beta_{0}+\beta_{1} T^{-1},
\end{equation}
where the coefficients $\beta_{0}$ and $\beta_{1}$ are
found by the least-square technique.
Formula (\ref{104}) is employed as a phenomenological
relation that describes the effect of temperature on
the segment length $b_{0}=b^{2}/L$.
Figure 7 reveals that (i) the number of statistical segments
increases with molecular weight,
and (ii) given $M_{\rm w}$, the Kuhn length decreases with
temperature (the growth of $T$ is tantamount to an increase
in the intensity of thermal fluctuations, which, in turn,
implies that the characteristic length of correlations
between monomers diminishes).

\section{Conclusions}

Introducing regularizing function $\rho$ in the form (\ref{5}),
we have derived explicit expressions for the scattering function
$S({\bf q})$ of a self-avoiding chain in a $d$-dimensional
space.
Analytical relations are developed for the form factor
and the radius of gyration of a self-repellent
chain in the one-dimensional and three-dimensional cases.
Results of numerical simulation demonstrate that the conventional
linear approximation for the radius of gyration (weak
excluded-volume interactions between segments) ensures an
acceptable agreement with the exact solution when the
dimensionless strength of interactions is less than unity.
Fitting of experimental data (small-angle neutron scattering)
on several polymer solutions reveals that the values of
the dimensionless strength of self-repellent interactions
exceeds the ultimate value at which this approximation is
reliable.

A novel approximation for the scattering function is proposed
(high strength of segment interactions $v\to \infty$
and large characteristic length of internal inhomogeneity
$L_{0}/L\to \infty$).
In the limit (\ref{100}), the scattering function
for a self-repellent chain in a three-dimensional space
is similar to that for a self-avoiding chain in an one-dimensional
space.
Equation (\ref{102}) deduced in this approximation provides
a simple generalization of the Debye formula (\ref{46}) for the
form factor of a standard Gaussian chain.
Comparison of numerical results with observations reveals
that Eq. (\ref{103}) adequately matches experimental data,
and its material parameters are affected by external conditions
in a physically plausible way.

\end{document}